\documentclass[aps,prd, nofootinbib, preprintnumbers, superscriptaddress, reprint]{revtex4-2}
\pdfoutput=1
\usepackage{amssymb, amsmath}
\usepackage{graphicx}
\usepackage{wrapfig}
\usepackage[normalem]{ulem}
\usepackage[colorlinks,linktocpage]{hyperref}
\usepackage{pict2e}

\usepackage{iftex}
\ifpdftex
\newcommand{\mbs}[1]{\boldsymbol{#1}}
\else
\usepackage[book]{fontsetup}
\newcommand{\mbs}[1]{\symbfit{#1}}
\fi

\hypersetup{
	colorlinks=true,
	citecolor=blue
}
\usepackage{xcolor}

\makeatletter
\@ifclasswith{revtex4-2}{preprint}
{
	\usepackage{setspace}
	\doublespacing
}
\makeatother

\newcommand{\vf}{\varphi}
\newcommand{\prt}{\partial}

\newcommand{\cb}{\mathrm{cb}}
\newcommand{\opn}[1]{\operatorname{#1}}

\usepackage[dvipsnames]{xcolor}

\begin{document}
	\preprint{INR-TH-2026-009}
	\title{Thermal false vacuum decay near black holes is aspherical}
	\author{D.S.~Gorbunov}
	\email{gorby@ms2.inr.ac.ru}
	\affiliation{Institute for Nuclear Research of the
		Russian Academy of Sciences, Moscow 117312, Russia}
        \affiliation{Moscow Institute of Physics and Technology,
        Dolgoprudny 141700, Russia}
	\author{D.G.~Levkov}
	\thanks{Deceased}
	%\email{levkov@ms2.inr.ac.ru}
	\affiliation{Institute for Nuclear Research of the
		Russian Academy of Sciences, Moscow 117312, Russia}
	\affiliation{Institute for Theoretical and Mathematical
		Physics, MSU, Moscow 119991, Russia}
	\author{V.E.~Maslov}
	\email{vasilevgmaslov@ms2.inr.ac.ru}	
	\affiliation{Institute for Nuclear Research of the
		Russian Academy of Sciences, Moscow 117312, Russia}
	\affiliation{Faculty of Physics, MSU, Moscow 119991, Russia}
	\begin{abstract}
          We study decay of a scalar field false vacuum near a
          (3+1)-dimensional Schwarzschild black hole equilibrated at
          Hawking temperature with the environment. Our scalar
          field model has negative  quartic self-coupling and thereby
          resembles Higgs sector of the Standard Model in the
          large-field limit. We demonstrate that if the black hole is
          not too small, the false vacuum in this model decays aspherically
          with regard to the black hole center: via formation of
          expanding true vacuum bubbles emerging on the 
          outer side of the
            event horizon. More
            specifically, we identify three regimes of the decay. For
            the largest and coldest black holes, the main mechanism is
            quantum tunneling described by an infinitesimally thin
            bounce sitting at some point of the  horizon. In the
            intermediate-mass regime, the vacuum is destroyed by
            thermal fluctuations creating aspherical critical bubbles
            in the horizon vicinity. Finally, near the smallest
            black holes thermal fluctuations still guide the decay
            but the dominant critical bubble is spherically symmetric
            and covers the entire horizon.
	\end{abstract}
	
	\maketitle
	
	\makeatletter
	\@ifclasswith{revtex4-2}{preprint}
	{
		\newpage
	}
	\makeatother

\section{Introduction and the main result}
It is well-known~\cite{Isidori:2001bm, Isidori:2007vm,
  Bednyakov:2015sca, DiLuzio:2015iua, Bednyakov:2025uur} that in
  the Standard model of particle physics the
  Electroweak vacuum is metastable (false) and can decay to subplanckian
  values of the Higgs field. In conditions of the present-day Universe
  the probability of such catastrophic event is safely
  small~\cite{Bednyakov:2025uur}. But there are proposals to catalyze
  the decay~--- notably, by considering it near black 
  holes (BHs): either isolated~\cite{Hiscock:1987hn, Berezin:1990qs,
    Gregory:2013hja, Burda:2015isa, Burda:2015yfa, Burda:2016mou} or
  in thermal equilibrium with the environment~\cite{Tetradis:2016vqb,
    Briaud:2022few}. Since the false Higgs vacuum has survived to date,
  the catalyzed processes, whenever efficient~\cite{Gorbunov:2017fhq,
    Mukaida:2017bgd, Hayashi:2020ocn, Shkerin:2021zbf, 
    Shkerin:2021rhy, Strumia:2022jil}, constrain cosmological
  scenarios with primordial 
  BHs and hot plasma~\cite{Gorbunov:2017fhq, Kohri:2017ybt,
    Hamaide:2023ayu} as well as models of particle
  physics~\cite{Dai:2019eei, Cuspinera:2019jwt}.

\begin{figure}
	  \includegraphics[width=8cm]{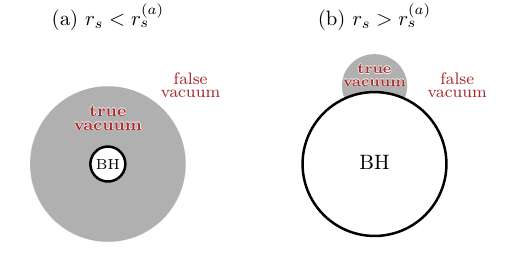}
	  \caption{Diagrams of bubbles formed during
              false vacuum decay near (a)~small and (b)~large
              Schwarzschild black holes of radius~$r_s$.}
          \label{fig:decay_schematic}
	\end{figure}

In this paper we study the false vacuum decay by a Higgs-like scalar
  field in the vicinity of a Schwarzschild black hole equilibrated at
  Hawking temperature\footnote{We use natural units with $\hbar =
  c = k_B = 1$.}~\cite{Hartle:1976tp, Gibbons:1976ue} 
\begin{equation}
  \label{eq:T_H}
  T = (4\pi r_s)^{-1}\,,
\end{equation}
with the surrounding gas, where~$r_s$ is BH radius.
We ignore backreaction of the scalar
  field on the spacetime metric. We show that~--- counter-intuitively
and against assumptions of the previous studies\footnote{Besides
BH-centered transitions, competing flat-space decays far away from the
BH are also considered~\cite{Shkerin:2021zbf, Miyachi:2021bwd}. We
argue that none of these processes are relevant in our realistic
(3+1)-dimensional model if the BH size is sufficiently  
  large.}~\cite{Tetradis:2016vqb, Briaud:2022few}~---  this
BH-induced decay is essentially aspherical with respect
to the BH center if the latter object is not too small. Namely,
the decay unsurprisingly creates true vacuum bubble filled with the
``subplanckian'' value of the  Higgs-like scalar field, see the gray
region in Fig.~\ref{fig:decay_schematic}. After formation the bubble
expands to engulf the Universe. But we show that the most probable
bubble looks like a spherically-symmetric ball in
  Fig.~\ref{fig:decay_schematic}(a) covering the entire event horizon 
  only if  the radius of the latter is smaller than some critical
  value~$r_s^{(a)} \propto m^{-1}$, 
  where~$m$ is the scalar field mass. At~$r_s >  r_s^{(a)}$ the
  spherical symmetry is spontaneously broken: the expanding bubble 
  emerges small  and squeezed to one side of the horizon, like in
  Fig.~\ref{fig:decay_schematic}(b). We believe that this qualitative
  picture is generic and inherent in all processes of false vacuum 
  decay near black holes, cf.~\cite{Berezin:1987ea, Arnold:1989cq,
    Berezin:1990qs, Gregory:2013hja, Burda:2015isa, Burda:2015yfa,
    Burda:2016mou, Mukaida:2017bgd, Hayashi:2020ocn}.

\begin{figure}[ht]
	  \centerline{\includegraphics{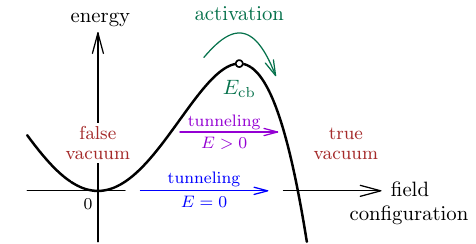}}
	  \caption{(Schematic) Configuration space and mechanisms
              for finite-temperature false vacuum decay.}
	  \label{fig:tunneling_ways}
	\end{figure}

        Specifically, we consider a (3+1)-dimensional model of 
          real scalar field~$\varphi(x)$ with mass~$m$ and potential
	\begin{equation}
		\label{eq:pot}
		V(\varphi) = \frac{m^2 \vf^2}{2} - \frac{\lambda \vf^4}{4}\,.
	\end{equation}
        Like the Higgs sector of the Standard Model at large 
        values of the 
          fields, this theory has negative quartic
          self-coupling~${-\lambda <  0}$ and two vacua:
          ``electroweak'' (false) at $\varphi = 0$ and ``subplanckian''
          (true) at $\varphi \to \infty$. The vacua are separated by
          an energy barrier of height~$E_{\mathrm{cb}}$, see
          the configuration space sketched in
          Fig.~\ref{fig:tunneling_ways}.

        In the absence of a black hole, one envisions three
          generic mechanisms~\cite{Kuznetsov:1997sf} of
          finite-temperature false vacuum decay in
          Fig.~\ref{fig:tunneling_ways}. First, the field can tunnel 
          through the barrier directly~\cite{Kobzarev:1974cp,
            Coleman:1977py, Coleman:1978ae}, regardless of the
          temperature. Second, 
          tunneling can happen at a positive energy taken from the
          thermal bath~\cite{Linde:1980tt, Linde:1981zj,
            Khlebnikov:1991th}. Third, the transition can be activated
          by thermal fluctuations pushing the field overbarrier
          classically~\cite{Langer:1969bc, Grigoriev:1989je, 
            Pirvu:2024nbe}. In all cases the decay rate is
          exponentially small,
          \begin{equation}
		\label{eq:exp-S}
		\Gamma \sim \mathrm{e}^{-S_E}\,, 
	  \end{equation}
with suppression~$S_E \propto \lambda^{-1}$ inversely
proportional to the coupling constant~$\lambda\ll 1$. But the outcomes
of the processes are different. The activation creates critical
bubbles~\cite{Kobzarev:1974cp, Coleman:1978ae, Rubakov:2002fi}~---
static classical solutions with energy~$E_{\mathrm{cb}}$ ``sitting''
on the barrier top\footnote{Similar solutions in gauge theories
  are called sphalerons~\cite{Manton:1983nd, Klinkhamer:1984di}.},
see Fig.~\ref{fig:tunneling_ways}. These bubbles are 
unstable: once perturbed, they expand and fill the space with the true
vacuum. In contrast, the two tunneling processes create bubbles with
zero or positive energy on  the ``true vacuum'' side of the
barrier. These bubbles start to expand immediately after nucleation. In the
semiclassical approximation~\cite{Coleman:1977py, Callan:1977pt,
  Coleman:1978ae, Khlebnikov:1991th}, the exponent~$S_E =
S_E[\varphi]$ of the decay rate in Eq.~(\ref{eq:exp-S}) is given by 
Euclidean action calculated on various Euclidean classical
solutions~--- bounces, periodic (thermal) bounces, and critical
bubbles, respectively, for the above three mechanisms.

The black hole background adds up spatial
inhomogeneity~\cite{Hiscock:1987hn, 
    Berezin:1990qs, Gregory:2013hja} and mass
defect~\cite{Tetradis:2016vqb} which affect the transition
  mechanisms making them more or less relevant. 
As a consequence, the 
  properties and suppression~$S_E$ of the thermal decay depend
  nontrivially on the BH radius~$r_s \propto T_H^{-1}$. In this paper
  we show that spherical symmetry also breaks down near
  not-so-small 
  black holes.

\begin{figure}
	  \includegraphics{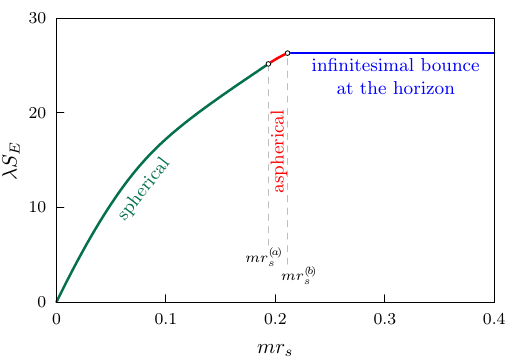}
	  \caption{Suppression exponent~$S_E$ of false vacuum
            decay near a Schwarzschild black hole of radius~$r_s$
              existing in thermal equilibrium with the
            environment. Model~(\ref{eq:pot}) is considered.}
          \label{fig:intro_answer}
	\end{figure}

The main result of our semiclassical study in the
  model~(\ref{eq:pot}) is shown in Fig.~\ref{fig:intro_answer}. We
  find that the false vacuum near the smallest and hottest black holes 
  decays via activation, i.e.\ by formation of static critical bubbles. The
  latter are
  \begin{subequations}  
  	  \label{eq:intervals_rs}  
    \begin{equation}
      \label{eq:small_bh}
      \mbox{spherical at} \quad  r_s \leq r_{s}^{(a)} \approx 0.194\,m^{-1}\,,
    \end{equation}
    like the one in Fig.~\ref{fig:decay_schematic}(a), and 
  \begin{equation}
    \label{eq:intermediate_bh}
    \mbox{aspherical at} \quad r_s^{(a)} \leq r_s \leq
      r_{s}^{(b)}\approx 0.211 \, m^{-1}
  \end{equation}
like the bubble in Fig.~\ref{fig:decay_schematic}(b). These two
regimes correspond to  
black holes with small, ${M_{\mathrm{BH}} \lesssim 0.097 \,
  M_{pl}^2/m}$, and intermediate masses, respectively. At even larger
black hole masses the dominant mechanism changes to 
  \begin{equation}
    \mbox{vacuum tunneling at} \quad r_s \geq r_{s}^{(b)} \,.
    \label{eq:lip_dom}
  \end{equation}
  \end{subequations}
  But unlike in the flat spacetime, the most probable transitions of this
  kind are also aspherical: they are described by extremely tall and narrow
  bounces~--- Fubini-Lipatov instantons (Lipatons)~\cite{Fubini:1976jm,
    Lipatov:1976ny, Affleck:1980mp, Linde:1981zj}~--- located at some
  point of the black hole event horizon. All of the arguments
    above support the expectation, that the transition is 
  aspherical at~${r_s > r_s^{(a)}}$.
        
The paper is organized as follows. In Sec.~\ref{Sec:Flat} we
review flat-space false vacuum decay at finite temperature
in the model~\eqref{eq:pot}. 
Adding a very large black hole in
  Sec.~\ref{Sec:Spher_modes}, we analytically prove that the activated
  decays near this object cannot be spherically-symmetric 
  with respect to its center. In Sec.~\ref{Sec:Aspherical} we numerically study
  the aspherical activation discriminating between dominant and
  subdominant critical bubbles. Tunneling via Fubini-Lipatov
  instantons is considered in Sec.~\ref{Sec:Lipatons}. In
Sec.~\ref{Sec:Discussion} we discuss results and future
applications. Analytic and numerical 
technicalities are consigned to Appendices.

%%%%%%%%%%%%%%%%%%%%%%%%%%%%%%%%%%%%%%%%%%%%

\section{False vacuum decay in the flat space}
\label{Sec:Flat}

To warm up, we review the thermal false vacuum decay in flat space
  filled with $\varphi$-quanta at temperature~$T$.  Thus we
    introduce
  relevant decay mechanisms in our (3+1)-dimensional scalar
  field model with potential~(\ref{eq:pot}) and Euclidean action
  \begin{equation}
  \label{eq:flat_action}
  S_E[\varphi] = \int d\tau  d^3 \mbs{x} \,\sqrt{g_E}
  \left[\frac{1}{2}\, g^{\mu\nu}_E \prt_{\mu} \vf \prt_{\nu}
    \vf  + V(\varphi)\right]\,, 
\end{equation}
where~$\tau \equiv it$ is Euclidean time and~$g^{\mu\nu}_E =
\delta^{\mu\nu}$ is flat metric. Note that the coupling
constant~${\lambda \ll 1}$ of the model governs semiclassical
expansion: after rescaling~$\vf \to \vf/\sqrt{\lambda}$ it enters
only as the overall prefactor~$\lambda^{-1}$ of the
classical action, at the place of~$\hbar^{-1}$ in the path 
integral.
	
In the zero-temperature case, the only way to permeate the
  potential barrier in Fig.~\ref{fig:tunneling_ways} is the quantum
  tunneling at~${E=0}$. In a semiclassical approximation this process 
  is described by the bounce, which is a time-dependent Euclidean solution with specific
  boundary conditions~\cite{Coleman:1977py, Coleman:1978ae,
    Rubakov:2002fi}. Euclidean action, calculated on the bounce
    configuration, $S_E[\varphi_b]$,  
  gives leading exponent of the decay rate~(\ref{eq:exp-S}). In the
  model~\eqref{eq:pot} with~${m>0}$ this solution is infinitesimally 
  small and has infinitely large field inside~\cite{Affleck:1980mp,
    Kuznetsov:1997az, Rubakov:2002fi}. Indeed, let us treat the mass
  term in the potential~(\ref{eq:pot}) as a perturbation. Then the
  leading-order theory is massless with quartic
  self-interaction~${-\lambda  \varphi^4/4}$. Relevant bounces in this
  case are Fubini--Lipatov instantons \cite{Fubini:1976jm, 
    Lipatov:1976ny} with arbitrary sizes~$a$,
  \begin{equation}
    \varphi_b(\tau, \mbs{x}) = \sqrt{\frac{8}{\lambda}}\,
    \frac{a}{a^2+\tau^2 + \mbs{x}^2} \qquad \text{at} \quad m=0\,.
    \label{eq:liprofile}
  \end{equation}
They take the form of~$O(4)$-invariant bubbles with radius~$a$, 
fields that scale as ${\varphi_b \propto a^{-1}}$ in the center, and identical
Euclidean actions
\begin{equation}
  S_E[\varphi_b]   = S_b \equiv \frac{8\pi^2}{3\lambda}  \qquad
  \mbox{at} \quad m=0
  \label{eq:lipaton}
\end{equation}
that yield the decay rate~(\ref{eq:exp-S}). One can argue
that~\cite{Coleman:1978ae, Rubakov:2002fi} the configuration emerging
after tunneling is a spherically-symmetric bubble $\varphi_b(0, 
\mbs{x})$. In the physical space-time with ${t = -i\tau}$ it expands, filling the 
space with ${\mbox{large-}\varphi}$ vacuum.

The degeneracy of $a$-dependent bounces originates from exact
  scaling symmetry $\varphi \to \varsigma \varphi(\varsigma  \tau,
  \varsigma \mbs{x})$ of the massless~$\varphi^4$ theory. Mass corrections 
  break this symmetry perturbing the bounces~\eqref{eq:liprofile} and
  adding $a$-dependent term to their actions~\cite{Affleck:1980mp, 
    Rubakov:2002fi},
  \begin{equation}
    \label{eq:m-bounce}
    S_E[\varphi_b]/S_b = 1 + c_b (ma)^2 + O(ma)^4\,,
  \end{equation}
  where~$S_b$ is given by Eq.~(\ref{eq:lipaton}) and numerical
    coefficient $c_b>0$ is 
  evaluated in Appendix~\ref{App:ReguLip} for completeness.
As a consequence, at~${m\ne 0}$  the only true saddle point of the
  Euclidean action is the configuration~(\ref{eq:liprofile})
  with~${a=0}$ and the smallest suppression $S[\varphi_b] = S_b$. One
  can regard these mass--perturbed configurations
  as constrained instantons~\cite{Affleck:1980mp,
  Rubakov:2002fi}, i.e. saddle points of $S_E$ at a fixed size $a$, and restore the true solution in the limit~$a\to 0$.

\begin{figure}[ht]
  \includegraphics{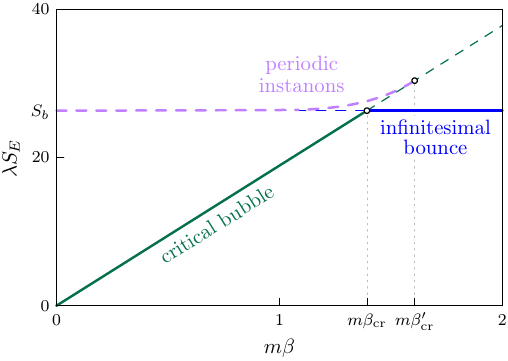}
  \caption{Thermal false vacuum decay in flat space:
    suppressions $S_E = S_E[\varphi]$ of various Euclidean solutions
    in the model~\eqref{eq:pot} as functions of 
    the inverse temperature ${\beta \equiv T^{-1}}$. Solid
    and dashed lines indicate dominant and exponentially
    suppressed contributions, respectively. The transition
    between tunneling via bounce and activation jumps
    over the critical bubble occurs at
    temperature $T_{\mathrm{cr}} \equiv \beta_{\mathrm{cr}}^{-1}$
    given in
    Eq.~\eqref{eq:Tcr-flat}.}
        \label{fig:flat_minphi4}
\end{figure}

Now, turn on nonzero temperature $T \equiv \beta^{-1}\ne
  0$. Since equilibrium quantum theory is~${\tau \to \tau +
  \beta}$ periodic in Euclidean time~\cite{Kapusta:2006pm}, 
  semiclassical solutions describing thermal transitions are 
  periodic as well. One of them can be
  constructed~\cite{Khlebnikov:1991th} by periodically placing
  infinitesimally thin bounces,  
  \begin{equation*}
    \varphi_{b}\Big|_{T\ne 0} = \sum_{n = -\infty}^{+\infty}
    \varphi_b(\tau+\beta n,\, \mbs{x}) \qquad \mbox{at}\quad  a\to 0\,.
  \end{equation*}
  This chain obeys Euclidean field equations at~${a\to 0}$ and has
  temperature-independent suppression~$S_E[\varphi_b] = S_b$ given by
  one-period Euclidean action~\cite{Linde:1980tt, Linde:1981zj,
    Khlebnikov:1991th} at~${|\tau|\leq \beta/2}$, see
  horizontal line in Fig.~\ref{fig:flat_minphi4}. Physically, the
  chain of infinitesimally thin bounces describes tunneling that
  happens at~$E=0$ despite thermal environment.

The second solution is time-independent, $\varphi=
  \varphi_{\mathrm{cb}}(\mbs{x})$, and hence $\tau$-periodic with
  arbitrary~$\beta$. It is represented by critical
  bubble~\cite{Kobzarev:1974cp, Coleman:1978ae, Rubakov:2002fi},
  that is a
  static configuration ``sitting'' on top of the potential barrier
  between the vacua in Fig.~\ref{fig:tunneling_ways}. In the
  model~(\ref{eq:pot}) this solution can be obtained
  numerically~\cite{Kuznetsov:1995cm, Kuznetsov:1997az} by integrating 
  the static field equation in the spherically-symmetric~\cite{Coleman:1977th}
  case~$\varphi_{\mathrm{cb}} = \varphi_{\mathrm{cb}}(|\mbs{x}|)$. One
  thus gets energy\footnote{Computed 
    as $E[\varphi]  \equiv \int d^3 x \, \left[\frac12 (\partial_t
      \varphi)^2 + \frac12 (\nabla_{\mbs{x}} \varphi)^2 + V(\varphi)
      \right]$.} and one-period Euclidean action,
  \begin{equation}
    \label{eq:Ecb}
    E_{\cb}  \approx 18.9 \; m/\lambda \qquad \mbox{and}\qquad
    S_E[\varphi_{\mathrm{cb}}] = \beta  
    E_{\mathrm{cb}}\,,
  \end{equation}
  of the critical bubble.

It is clear that $\varphi_{\mathrm{cb}}(\mbs{x})$ describes
 the activation, that is the overbarrier transitions with energy $E_{\mathrm{cb}}$
  sourced by fluctuations of the thermal bath, see
  Fig.~\ref{fig:tunneling_ways}. Indeed, this 
  configuration literally sits on the barrier top and gives
  Boltzmann-suppressed rate~(\ref{eq:exp-S}), 
  \begin{equation}
    \label{eq:exp_bubble}
    \Gamma \propto \mathrm{e}^{-S_E[\varphi_{\mathrm{cb}}]} =
    \mathrm{e}^{-E_{\mathrm{cb}}/T}\,,
  \end{equation}
  see Eq.~(\ref{eq:Ecb}). The suppression exponent of this
  solution is shown by solid and dashed diagonal lines in
  Fig.~\ref{fig:flat_minphi4}.

Finally, one can numerically compute~\cite{Kuznetsov:1997az,
    Kuznetsov:1997sf} periodic instantons~\cite{Linde:1980tt,
    Linde:1981zj, Khlebnikov:1991th, Kuznetsov:1997sf}, which are 
  $\tau$-dependent genuinely smooth solutions~${\varphi =
    \varphi_{\mathrm{PI}}(\tau,\, \mbs{x})}$ with
  $\tau$-periods~$\beta$.\footnote{Note that periodic instantons
    have turning points $\partial_\tau \varphi = 0$ at ${\tau= 0}$
    and~$\beta/2$~\cite{Khlebnikov:1991th, Kuznetsov:1997az}, and
      hence 
     they can be computed by solving the Euclidean field equation
    with Neumann boundary conditions.} They describe tunneling at
  nonzero energies ${E = E_{\mathrm{PI}}(\beta)}$ taken from the
  thermal bath, see Fig.~\ref{fig:tunneling_ways}. The
    suppression 
  exponent~$S_E[\varphi_{\mathrm{PI}}]$ of periodic instantons is
  depicted on the plot of Fig.~\ref{fig:flat_minphi4} as a dashed curve. Notably,
  they disappear below certain temperature~${T'_{\mathrm{cr}} \equiv
  (\beta_{cr}')^{-1}}$ by becoming $\tau$-independent and turning into
  critical bubbles.

At the culmination of this Section we identify physically relevant
  mechanisms of false vacuum decay in the
  model~(\ref{eq:flat_action}), (\ref{eq:pot}). Since we are working
  in the leading semiclassical order, it is sufficient to select the
  contribution with the smallest suppression~$S_E$ at every
  temperature. This implies that the infinitesimal
  bounce~$\varphi_b$ and critical 
  bubble~$\varphi_{\mathrm{cb}}$ are relevant only at low   
  and high temperatures, respectively, see two solid lines in
  Fig.~\ref{fig:flat_minphi4}, whereas the entire branch of periodic  
  instantons~$\varphi_{\mathrm{PI}}$ (dashed curve) is unphysical.
  Transition between  the regimes of vacuum tunneling
  via~$\varphi_b$ and activation via~$\varphi_{\mathrm{cb}}$ happens
  at critical temperature  
  \begin{equation}
    \label{eq:Tcr-flat}
    T_{\mathrm{cr}} \equiv \beta_{\mathrm{cr}}^{-1} =
    E_{\cb}/S_b  \approx 0.718 \,  m 
  \end{equation}
  when the two suppressions coincide\footnote{At~$T\approx
    T_{\mathrm{cr}}$ one is tempted to include both equally suppressed
    contributions. But that would require calculation of prefactors
    and proper resurgent analysis which is beyond the scope of this
    paper, cf.~\cite{Hampton:2026uap}.}, $S_b = E_{\cb} / T_{\mathrm{cr}}$, see
  Eqs.~(\ref{eq:lipaton}) and (\ref{eq:Ecb}).

Below we extend this intuition to the problem of thermal
  false vacuum decay near black holes. Namely, we assume that
  finite-energy periodic instantons are subdominant in that case as
  well and study only critical bubbles and infinitesimal
  instantons. This strategy is supported by the
  literature~\cite{Arnold:1989cq, Briaud:2022few, Shkerin:2024igk} 
  suggesting that periodic instanton contributions are irrelevent near black
  holes, indeed.

It is worth noting that the dominant mechanism of  thermal
  transitions can be selected on solid theoretical 
  grounds~\cite{Khlebnikov:1991th, Kuznetsov:1997sf}. To this end one 
  notes~\cite{Khlebnikov:1991th} that every single solution
  considered in this Section is physical, albeit they all describe
  decays of fixed-energy excited states above the false vacuum
    rather than thermal decays, with the rate
  \begin{equation}
    \label{eq:P_E}
    \Gamma_E \sim \mathrm{e}^{- S_E[\varphi] + E \beta_E} \qquad
    \mbox{at} \quad E \leq E_{\mathrm{cb}} 
  \end{equation}
  and $\Gamma_{E} \sim 1$ at $E>E_{\mathrm{cb}}$,  where~$\beta_E$ is a
  Euclidean period depending on the energy~$E$. In particular, infinitesimal 
  bounce corresponds to tunneling at~$E = 0$, critical bubble
  describes transition at $E = E_{\mathrm{cb}}$, while periodic
  instantons work at intermediate energies. Thermal
  rate~(\ref{eq:exp-S}) is obtained by averaging Eq.~\eqref{eq:P_E}
  over canonical ensemble in the false
  vacuum~\cite{Kuznetsov:1997sf},  
  $$
  \Gamma \propto \int dE\, \mathrm{e}^{-\beta E}\, \Gamma_E  \sim \int dE \,
 \mathrm{e}^{- S_E[\varphi] + E (\beta_E - \beta)}\,.
  $$
 To the leading semiclassical order, this formula means maximization
 of the integrand with respect to its boundary values at $E = 0$ and
   $E_{\mathrm{cb}}$ and possible extrema\footnote{Extremum
   at intermediate energies is achieved at $\beta_{E}  = \beta$
   because~\cite{Kuznetsov:1997sf} $\partial_E
   S_{E}[\varphi_{\mathrm{PI}}] = E\, \partial_E 
   \beta_E$ for the solution with
   period~$\beta_E$.} in between: $\Gamma  \propto  
  \max\{ \mathrm{e}^{-S_E[\varphi_b]},\, \mathrm{e}^{- \beta
    E_{\mathrm{cb}}},\,
  \mathrm{e}^{-S_E[\varphi_{\mathrm{PI}}]}\}$. This 
  explains the selection procedure in Fig.~\ref{fig:flat_minphi4}.	

%%%%%%%%%%%%%%%%%%%%%%%%%%%%%%%%%%%%%%%%%%%%%%%%%%

\section{Spherical critical bubbles turn unphysical}
\label{Sec:Spher_modes}

In this Section we study activated decay of false vacuum near
  nonrotating black holes of radius~$r_s$ equilibrated with thermal
  bath of $\varphi$-particles. We deliberately start from
  spherically-symmetric critical bubbles~\cite{Tetradis:2016vqb,
    Briaud:2022few} which are definitely physical at~${r_s \ll
    m^{-1}}$ when the Hawking temperature~\eqref{eq:T_H} is high and
  the black hole attraction is weak. We will argue, however, that the
  same spherical bubbles become irrelevant at  large~$r_s$~--- not due
  to changing transition mechanism, like in 
  flat space, but with hints towards aspherical transitions.

It is natural to assume that the black hole is much heavier than
  the critical bubbles,
  \begin{equation}
    \label{eq:BH_inequality}
    M_{\mathrm{BH}} \equiv \frac12 M_{pl}^2\, r_s \gg E_{\mathrm{cb}}\;.
  \end{equation}
  Indeed, exponentially suppressed activation occurs at
  ${E_{\mathrm{cb}} < T_H}$ implying
  $E_{\mathrm{cb}}/M_{\mathrm{BH}} < T_H/M_{\mathrm{BH}} \sim
  {M_{pl}^2/M_{\mathrm{BH}}^2 \ll 1}$, where we used the Hawking 
  temperature~\eqref{eq:T_H} and recalled that
  semiclassically tractable black holes are superplanckian.

Inequality~\eqref{eq:BH_inequality} means that we can neglect
  backreaction of the scalar field on the Schwarzschild metric
  which is therefore fixed,
\begin{equation}
  ds^2_E = \left(\frac{R - R_s}{R + R_s}\right)^2
  d\tau^2 + \frac{r^2(R)}{R^2} \left(dR^2 + R^2
    d\Omega^2\right)\,.
    \label{eq:isotropic_metric}
\end{equation}
Here the isotropic radius~$R$ is related~\cite{Weyl:1917gp} to
  the standard Schwarzschild coordinate~$r$ as
\begin{equation}
  r(R) = R \left(1+R_s/R\right)^2\,, \qquad R_s \equiv
  \frac{r_s}{4}\,,
  \label{eq:isotropic_r}
\end{equation}
we use  Euclidean time~$\tau \equiv it$ and spherical surface
  element $d\Omega^2 = d \theta^2 + \sin^2 \theta \, d
  \phi^2$. These coordinates are convenient for studying
  rotational asymmetry because spatial sections of the
  metric~\eqref{eq:isotropic_metric} are conformally flat.

\begin{figure}[ht]
  \centerline{\includegraphics{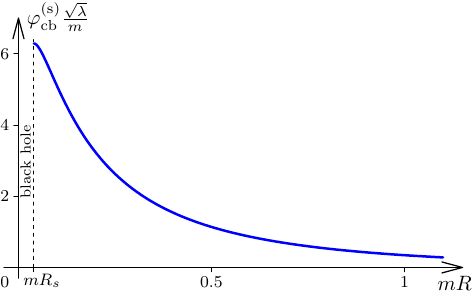}}
  \caption{Spherically-symmetric critical bubble $\varphi =
      \varphi_{\mathrm{cb}}^{\mathrm{(s)}} (R)$ in Schwarzschild
      background with~$r_s = 0.16\, m^{-1}$.}
  \label{fig:BH_cbs}
\end{figure}

\begin{figure}[ht]
  \includegraphics{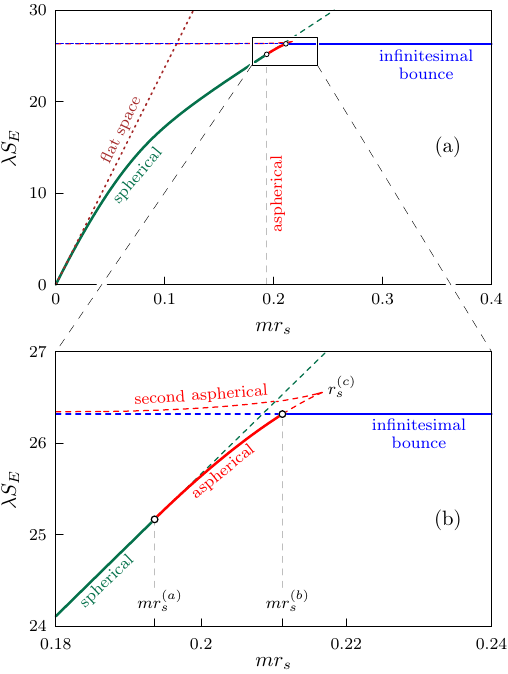}
  \caption{Euclidean actions~$S_E[\varphi]$ of all static true
      vacuum bubbles in Schwarzschild background. Solid and dashed
      lines mark dominant and subdominant contributions to thermal
      false vacuum decay, respectively. Figure~(b) zooms into the
    transition region in Fig.~(a) between spherical critical
    bubbles, aspherical critical bubbles and infinitesimal
    bounces.}
  \label{fig:full_res}
\end{figure}

Substituting rotationally-invariant Ansatz $\varphi =
  \varphi_{\mathrm{cb}}^{\mathrm{(s)}}(R)$ into the static field
  equation of the model~(\ref{eq:flat_action}),
  \eqref{eq:isotropic_metric}, we get an equation
  for the spherical bubble profile, 
  \begin{equation}
    \label{eq:sph}
    \partial^2_R \vf_{\mathrm{cb}}^{\mathrm{(s)}} + \frac{2R \;
      \partial_R \vf_{\mathrm{cb}}^{\mathrm{(s)}}}{R^2 - R^2_s} =
    \frac{r^2(R)}{R^2} \, V'(\vf_{\mathrm{cb}}^{\mathrm{(s)}}) \,,
  \end{equation}
  where~$V(\varphi)$ is given by Eq.~(\ref{eq:pot}) and the prime
  is its~$\varphi$ derivative. Equation~(\ref{eq:sph}) can be solved
  numerically by shooting~\cite{Press:2007ipz, Levkov:2017paj} with 
  regularity conditions at the horizon and at infinity,
  \begin{equation}
    \partial_R \varphi_{\mathrm{cb}}^{\mathrm{(s)}}(R_s) = 0
    \qquad \text{and} \qquad
    \varphi_{\mathrm{cb}}^{\mathrm{(s)}} \to 0 \; \; \text{as}\;\; R \to \infty\,,
    \label{eq:bc_sph}
  \end{equation}
  see Appendix~\ref{App:Modes} for their derivation. It produces smooth
  family of legitimately-looking critical bubbles parametrized
  with~$r_s$, see an example in
  Fig.~\ref{fig:BH_cbs}. Suppressions~$S_E[\varphi_{\mathrm{cb}}^{\mathrm{(s)}}]$
  of these solutions, Eq.~(\ref{eq:lipaton}), are displayed in
  Fig.~\ref{fig:full_res}  (green solid curve continued with the
  dashed one). They are
  lower~\cite{Tetradis:2016vqb}  than  flat-space suppressions
  (dashed straight line on the plot), which indicates that black
  holes do catalyze activated transitions.

Recall, however, that the space of static field
  configurations~$\varphi(\mbs{x})$ in Fig.~\ref{fig:tunneling_ways}
  is infinite-dimensional. The height~$E_{\mathrm{cb}} \equiv
  S_E[\varphi_{\mathrm{cb}}]/\beta$ of the barrier separating the
  vacua is defined as a minimum energy of classical trajectories
  interpolating between them. This means that the true critical
  bubble~$\varphi_{\mathrm{cb}}(\mbs{x})$ with~${E = E_{\mathrm{cb}}}$
  occupies the lowest saddle point of~$S_E[\varphi]$ in the
  barrier region. It has precisely one negative
  mode~\cite{Coleman:1987rm, Briaud:2022few}~---  direction downhill
  leading to the vacua~--- whereas other changes
  of~$\varphi_{\mathrm{cb}}(\mbs{x})$ should increase~$S_E$ 
  and suppress  the activation rate~(\ref{eq:exp-S}).  

To count negative modes of the spherically-symmetric
bubble~$\varphi_{\mathrm{cb}}^{\mathrm{(s)}}$, we add a small static
variation~$\delta \varphi$,  
\begin{equation}
  \label{eq:11}
  \varphi = \varphi_{\mathrm{cb}}^{\mathrm{(s)}} (R) +
  \delta \varphi(\mbs{x})\,.
\end{equation}
The latter can be naturally decomposed in the basis of 
eigenmodes~$\xi_{k\ell}$ diagonalizing the second variation of the
classical action~(\ref{eq:flat_action}) above
$\varphi_{\mathrm{cb}}^{\mathrm{(s)}}$:
\begin{equation}
  \delta \varphi = \sum\limits_{k\ell m}A_{k\ell m} \, \xi_{k\ell}(R) \, Y_{\ell
    m} (\theta, \phi) \,,
  \label{eq:sph_decomp}
\end{equation}
so that\footnote{We use normalization specified in
  Appendix~\ref{App:Modes}.} to the quadratic order in the field
  variations 
\begin{equation}
  \label{eq:13}
  S_E \approx
  S_E[\varphi_{\mathrm{cb}}^{\mathrm{(s)}}] + 
  \frac12 \sum\limits_{k\ell m}\, \mu_{k\ell}\, |A_{k\ell m}|^2 \,,
\end{equation}
where we introduced radial~$k$ and rotational~$\ell$, $m$ quantum 
numbers, an eigenvalue~$\mu_{k\ell}$ of~$\xi_{k\ell}$, as well as spherical
harmonics~$Y_{\ell m}$.  Now, it is transparent that negative modes
which decrease the action have~$\mu_{k\ell} <0$.

An eigenproblem for the modes~$\xi_{k\ell}$,
\begin{subequations}
  \label{eq:eigen_combo}
  \begin{equation}
    \label{eq:eigen}
    \hat{L}_{\ell}  \xi_{k\ell} (R)= \mu_{\ell} \xi_{k\ell}(R)\,,
  \end{equation}
follows from the action~(\ref{eq:flat_action}) in the background
(\ref{eq:isotropic_metric}), see Appendix~\ref{App:Modes} for
derivation. Specifically, the operator  
\begin{gather}
  \label{eq:rad_operator} 
  \hat{L}_{\ell} = -\partial^2_R -   \frac{2R\, \partial_R}{R^2 -
    R^2_s}   + \frac{\ell(\ell+1)}{R^2} + \frac{r^2}{R^2}\,
  V''(\varphi_{\mathrm{cr}}^{\mathrm{(s)}}) 
\end{gather}
can be read off the second variation of~$S_E$ around
$\varphi_{\mathrm{cr}}^{\mathrm{(s)}}$, whereas the regularity
conditions 
\begin{equation}
  \prt_R \xi_{k\ell}(R_s) = 0 \qquad
  \text{and} \qquad \xi_{k\ell} \to 0\; \text{as} \; R \to
  \infty\,,
  \label{eq:bc_mod}
\end{equation}
\end{subequations}
are the same as in Eq.~(\ref{eq:bc_sph}).

Unfortunately, Eqs.~(\ref{eq:eigen_combo}) cannot be solved
  analytically, given that the background~$\varphi_{\mathrm{cb}}^{\mathrm{(s)}}(R)$ is
  numerical. Let us, however, give a convincing argument that
  spherically-symmetric critical  bubbles are physical at~$mr_s \ll 1$ 
  and unphysical at~$mr_s \gg 1$.

At $mR_s \sim mr_s  \to 0$ the profile and eigenmode
  equations~(\ref{eq:sph}) and~(\ref{eq:eigen_combo}) coincide with
  their flat-space limits:  the black hole is negligibly small. We
  already mentioned in Sec.~\ref{Sec:Flat} that flat-space
  critical bubbles in the model~(\ref{eq:pot}) are spherical and
  physical. This implies that they have precisely one 
  negative mode with~$\ell=0$, see Ref.\,\cite{Kuznetsov:1997az}. 

In the opposite limit~$mr_s \to +\infty$ the thickness~$m^{-1}$ of
  the bubble walls is small compared to the black hole
  radius~$R_s$. This makes spherical bubble resemble thin shell
  located at~$R\approx R_s$.  Taking the near-horizon limit $\rho = R -   
  R_s \ll R_s$ in the profile equation~\eqref{eq:sph}, we find,
  \begin{equation}
    \prt^2_{\rho}
    \vf_{\mathrm{cb}}^{\mathrm{(s)}} + \frac{\prt_{\rho} 
      \vf_{\mathrm{cb}}^{\mathrm{(s)}}}{\rho} = 16
    V'(\vf_{\mathrm{cb}}^{\mathrm{(s)}})\,, \qquad mr_{s} \gg 1\,.
    \label{eq:2d_bubble}
  \end{equation}
  It is remarkable that this limiting equation has the same form as
  the one for rotationally-invariant critical
  bubble~$\vf_{\mathrm{cb}}^{\mathrm{(s)}}(\rho)$ in the fictitious 
  two-dimensional flat-space theory with potential~$16 V(\varphi)$,
  radial coordinate $\rho \geq 0$, and radial Laplacian  
  $\Delta_{2d} \equiv  \partial_\rho^2 + \rho^{-1}
  \partial_\rho$. Moreover, the eigenvalue
  problem~\eqref{eq:eigen_combo} at~${R \approx R_s}$ reduces to
  equation
  \begin{equation}
    \label{eq:rad_modes_limit}
    -\prt^2_{\rho} \xi_{k\ell} - \frac{\prt_{\rho} \xi_{k\ell}}{\rho} +
    16 V''(\varphi_{\mathrm{cb}}^{\mathrm{(s)}}) \xi_{k\ell} =
    \mu_{2d} \,\xi_{k\ell}\,,
  \end{equation}
  for nonrotating modes of the critical bubble in the same fictitious
  theory, and we introduced $\mu_{2d} \equiv \mu_{k\ell} -
  \ell(\ell+1)/R^2_s$ that impersonates the mode eigenvalue in two dimensions.

  It is important to keep in mind, however, that the near-horizon limit of
  our model is not equivalent to flat two-dimensional
  theory. In truth,~$\rho=0$ is a position of the spherical horizon, not an origin of
  space. Hence, modes with nonzero~$\ell$ and~$m$ behave
  differently in three dimensions: at $mr_s \gg 1$ they satisfy one
  and the same ``nonrotating'' equation~(\ref{eq:rad_modes_limit}),
  albeit  with shifted~$\mu_{2d}$.

  And this property makes the two pictures fundamentally different. In
  flat two-dimensional space, the critical bubble obtained from 
  Eq.~(\ref{eq:2d_bubble}) is physical. It describes the thermal
  activation and has precisely one nonrotating negative mode 
  satisfying Eq.~(\ref{eq:rad_modes_limit}). Numerical
  integration~\cite{Press:2007ipz, Levkov:2017paj} of
  Eqs.~(\ref{eq:2d_bubble}), (\ref{eq:rad_modes_limit}) confirms this
  expectation and provides the eigenvalue 
  \begin{equation}
    \label{eq:mu_2d}
    \mu_{2d}\approx - 86.6\, m^{2}\,.
  \end{equation}
  Expressing~$\mu_{2d}$ via~$\mu_{k\ell}$, we reformulate the same
  result in the original model: all sectors with
  \begin{equation}
    \label{eq:number_of_modes}
    \ell \leq \ell_{\mathrm{max}}\equiv \mathcal{I}\left[\sqrt{|\mu_{2d}| R^2_s +
	\tfrac{1}{4}} - \tfrac{1}{2}\right] \,, \qquad mr_s \gg 1\,,
  \end{equation}
  and arbitrary $m$ have one negative mode~$\mu_{k\ell} < 0$ each,
  where ${\cal I}[A]$ is 
  an integer part of $A$. At $mr_s \gg 1$ the number ${N_{\opn{modes}} \equiv
  (\ell_{\mathrm{max}} + 1)^2 \gg 1}$ of such negative modes with 
  ${\ell \leq \ell_{\mathrm{max}}}$ is enormous. We 
  conclude that spherical critical bubbles surrounding black holes are
  physical with ${N_{\opn{modes}} =1}$  only if the latter objects are
  sufficiently small:~${r_s < r_s^{(a)} \sim m^{-1}}$. 

  To confirm the analytical arguments, we numerically count the number
  $N_{\mathrm{modes}} \equiv (\ell_{\mathrm{max}}+1)^2$ of negative modes
  in the full eigenvalue problem~(\ref{eq:eigen_combo}) using
  oscillation theorem, see Appendix~\ref{App:Modes} for
  details. In Fig.~\ref{fig:Nmodes_Rs} we present this number at
  different~$r_s$ (circles) together with the large-$r_s$
  prediction~\eqref{eq:mu_2d}, (\ref{eq:number_of_modes}) (stair-like  
  solid line). We see indeed that the critical bubble has a single
  negative mode with $\ell=0$ only at $r_s < r_s^{(a)}$,
  where~$r_s^{(a)}$ is  given by Eq.~(\ref{eq:small_bh}).  At larger
  $r_s$ three dipole modes with~$\ell=1$ become negative, too. The number
  of such modes grows with~$r_s$ and eventually approaches the 
  asymptotics~\eqref{eq:mu_2d},  (\ref{eq:number_of_modes}).  

\begin{figure}
  \centering
  \unitlength=1mm
  \begin{picture}(86,61)
    \put(0,0){\includegraphics{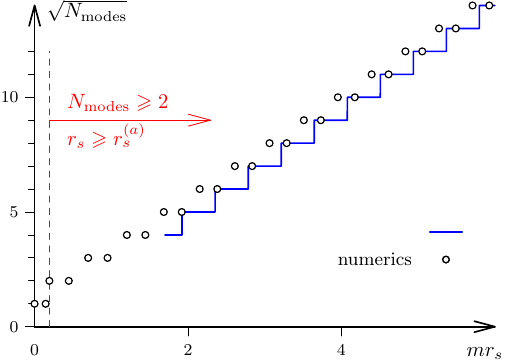}}
    \put(70,21.5){\makebox(0,0)[r]{
        asymptotics~\eqref{eq:mu_2d}, \eqref{eq:number_of_modes}}}
  \end{picture}
  \caption{Number of negative modes $N_{\mathrm{modes}} \equiv
    (\ell_{\max}+1)^2$  in the background of a
    spherically-symmetric critical bubble centered on the Schwarzschild
      black hole of radius~${r_s\equiv
        4R_s}$. Numerical results (circles) are
      compared to large-$r_s$
    asymptotics~\eqref{eq:mu_2d} \eqref{eq:number_of_modes} (line).} 
  \label{fig:Nmodes_Rs}
\end{figure}

  To sum up, spherical  critical bubbles become unphysical
    at~${r_{s} > r_s^{(a)}}$ because they acquire additional, 
    rotationally asymmetric, negative modes that
    decrease~$S_E$. Thus, dominant activated transitions are
    aspherical at~${r_s > r_s^{(a)}}$. On the contrary, we show in  
    Appendix~\ref{App:Modes} that time-dependent perturbations ${\delta
    \varphi = \delta\varphi(\tau, \mbs{x})}$ never
    decrease~$S_E$. This implies that time-dependent Euclidean
    solutions, periodic  instantons,  are physically irrelevant.

%%%%%%%%%%%%%%%%%%%%%%%%%%%%%%%%%%%%%%%%%%%%

\section{Aspherical bubbles}
\label{Sec:Aspherical}
  
Suppose the BH radius~$r_s$ is slightly above the
  threshold~$r_{s}^{(a)}$. This makes the spherical
  bubbles~$\varphi_{\mathrm{cb}}^{\mathrm{(s)}}$ unphysical  
  with four negative modes: dipole~$\xi_{01}$ with~${\ell = 1}$, 
  ${m=0,\, \pm 1}$ and monopole~$\xi_{00}$ 
  with~$\ell=0$.  It is natural to expect that the dominant critical
  bubble, solution with smaller~$S_E$  describing 
  	activation, differs from~$\varphi_{\mathrm{cb}}^{\mathrm{(s)}}$
  in the direction of these modes,
  \begin{equation}
    \vf = \varphi_{\mathrm{cb}}^{\mathrm{(s)}}(R) + A_0\, \xi_{00}(R) + A_1\, \xi_{01}(R)
    \cos \theta \label{eq:spher_deformed}\,,
  \end{equation}
  where we exploit decomposition~\eqref{eq:sph_decomp} with~${A_{\ell}  \equiv
    A_{0\ell0}}$ but take into account only
  axially symmetric negative modes with~${Y_{00} \propto 1}$
  and~${Y_{10} \propto \cos \theta}$. 

\begin{figure}
  \includegraphics[trim={0 0 0 5mm},clip]{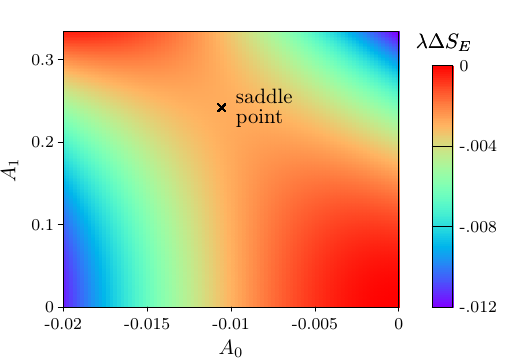}
  \caption{Euclidean action $S_E(A_0, A_1)$ of aspherical
    configurations~\eqref{eq:spher_deformed} depending on the 
    deformation parameters~$A_{\ell}$ at~$r_s = 0.2\,
      m^{-1}$. Color shows $\Delta S_E \equiv S_E(A_0, A_1) - S_E(0, 
    0)$, black cross marks saddle point at ${(A^{\!\times}_0,\,
      A^{\!\times}_1) \approx (-0.01056,\, 0.242)}$} 
  \label{fig:SE_A0_A1}
\end{figure}

  The configuration~(\ref{eq:spher_deformed}) satisfies static field
  equation up to second-order corrections in~$A_{\ell}$. Let
  us compute its one-period Euclidean action $S_E(A_0,
  A_1)$. Solving numerically Eqs.~\eqref{eq:sph} and
  (\ref{eq:eigen_combo}) for $\varphi_{\mathrm{cb}}^{\mathrm{(s)}}$,
  $\xi_{00}$, and~$\xi_{01}$, we calculate the integral
  (\ref{eq:flat_action}) which in static axially-symmetric 
  case simplifies to
  \begin{multline}
    S_E = 32\pi^2R_s \int d\theta dR \,  \sin \theta \, (R^2 -
    R^2_s) \times \\ \left[\frac{\left(\prt_R \vf\right)^2}{2} +
      \frac{\left(\prt_\theta \vf\right)^2}{2R^2} +
      \frac{r^2(R)}{R^2}\, V(\vf)\right]\,,
		\label{eq:eucl_action_asym}
  \end{multline}
  see the metric \eqref{eq:isotropic_metric} and
  	\eqref{eq:isotropic_r}
  	and Hawking period~${\beta = T^{-1}}$ in Eq. (\ref{eq:T_H}).
  Figure~\ref{fig:SE_A0_A1} shows the numerical
  result~--- function~$S_E(A_0, A_1)$ at~${r_s = 0.2\, m^{-1}} >
  r_s^{(a)}$. It has two\footnote{Plus mirror-symmetric  
  saddle point $(A_0, A_1) = (A^{\!\times}_0,
    -A^{\!\times}_1)$.} extrema: maximum at~${A_0 = A_1 = 0}$ (right 
  bottom corner) and a saddle  point $(A_0,A_1) = (A^{\!\times}_0,
  A^{\!\times}_1)$ with smaller~$S_E$ and a single negative mode (the
  cross). The latter saddle point approximately represents dominant
  critical bubble, that is the aspherical one resembling the
  configuration~\eqref{eq:spher_deformed}  with
  parameters~$A^{\!\times}_{\ell}$.  

  The dominant bubble is expected to be $\phi$-independent. First, we
  already saw that the deformation~(\ref{eq:spher_deformed}) leaves
  only one  
  negative mode in the axially-symmetric sector. Modes with~$\phi$
  dependence have strictly larger eigenvalues, as their eigenproblem
  includes additional positive-definite
  operator~$-\partial_\phi^2$. As a consequence, appearance of
  any~$\phi$-dependent eigenmode with  $\mu \leq 0$ guarantees
  existence of its axially-symmetric counterpart negative
  mode. Second,  there are two zero modes with~$\mu=0$ representing
  infinitesimal rotations of aspherical bubbles; they are proportional
  to $\mathrm{e}^{\pm i\phi}$. The counterpart of these modes in the
  axially-symmetric sector is the negative mode itself. Thus, there
  are no other $\mathrm{e}^{\pm i \phi}$ modes with $\mu \leq 0$
  because there are no  other counterparts.  We conclude that the
  deformation (\ref{eq:spher_deformed}) lifts the $\ell =1$ and $m = \pm
  1$ negative modes of $\varphi_{\mathrm{cb}}^{\mathrm{(s)}}$ to zero
  modes with~${\mu = 0}$. 

  We compute precise aspherical critical bubbles ${\varphi =
  \varphi_{\mathrm{cb}}^{\mathrm{(a)}}}$ numerically. To this end we
  impose axial symmetry around the axis~${\theta=0}$ connecting the 
  bubble and black hole centers, discretize the
  action~(\ref{eq:eucl_action_asym}) on the lattice covering a
  region of~$R$ and~$\theta$, then solve 
  finite-difference equations $\delta S_E/\delta \vf = 0$ for the
  saddle point~$\varphi_{\mathrm{cb}}^{\mathrm{(a)}}(R,\theta)$ using
  Newton--Raphson iterations~\cite{Demidov:2011dk,
    Demidov:2015bua, Demidov:2015nea,  Levkov:2017paj,
    Demidov:2022ljh} and conjugate gradients linear
  solver~\cite{Press:2007ipz},  see Appendix~\ref{App:Numerical} for  
  details and numerical tests. Our method converges to a full nonlinear
  solution but needs good initial approximation.

\begin{figure}
  \includegraphics[trim={0 0 0 1mm},clip]{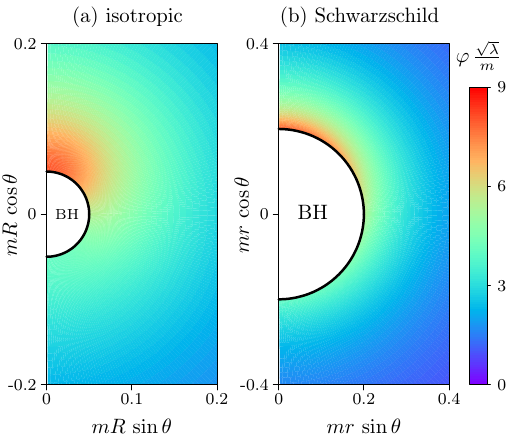}
  \caption{Aspherical critical bubble at $r_s \equiv
    4R_s = 0.2 \, m^{-1}$ in
    (a)~isotropic and (b)~Schwarzschild coordinates.}
  \label{fig:aspher_0.05}
\end{figure}

  In particular, we obtain the solution at ${r_s = 0.2\,
    m^{-1}}$ using the configuration~(\ref{eq:spher_deformed}) with
  parameters~$A_0^\times$ and~$A_1^\times$ as an initial seed. The
  resulting critical bubble is shown in Fig.~\ref{fig:aspher_0.05} in
  isotropic~$(R,\theta)$ and Schwarzschild $(r,\theta)$ coordinates
  [recall Eq.~(\ref{eq:isotropic_r})]. It is essentially anisotropic. 

  Once the first solution is found, we start changing the
  black hole radius~$r_s$ in small steps and getting a new numerical 
  solution at every step, with the previous solution serving as an initial
  approximation. This recovers a smooth branch of aspherical critical
  bubbles at $r_s^{(a)} < r_s < r_s^{(c)}$. Their Euclidean
  actions~$S_E[\varphi_{\mathrm{cb}}^{\mathrm{(a)}}]$ are shown in  
  Fig.~\ref{fig:full_res} by lower red lines marked
  ``aspherical'' (solid and dashed). These suppressions are smaller
  than in the spherical case, cf.\ the green dashed line. 

  We expect that each aspherical bubble in the interval
  ${r_s^{(a)} < r_s < r_s^{(c)}}$ has precisely one negative
  mode. Indeed, at~${r_s \approx r_s^{(a)}}$ we explicitly excluded
  excessive modes of this kind. Our subsequent numerical procedure
  was based on inverting the second variation of~$S_E$ in the
  background of solution at every $r_s$, cf.\ Eq.~(\ref{eq:eigen}) and
  Appendix~\ref{App:Numerical}. It 
  would diverge if some mode eigenvalue crossed zero to become
  negative.

\begin{figure}[ht]
  \centering
  \unitlength=1mm
  \begin{picture}(86,61)
    \put(0,0){ \includegraphics{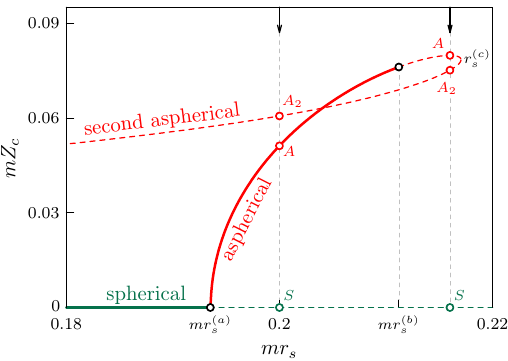}}
    \put(41.24,57){\scriptsize \ref{fig:branches}(a)}
    \put(70,57){\scriptsize \ref{fig:branches}(b)}
  \end{picture}
  \caption{Asphericities~$Z_c$ of critical bubbles versus 
      the BH radius~${r_s\equiv 4R_s}$. Line styles and colors are the
      same as in Fig.~\ref{fig:full_res}. Arrows show the values
    of~$r_s$ in Figs.~\ref{fig:branches}(a) and~(b), circles $A$,
      $A_2$, and~$S$ are the extrema in these figures.} 
  \label{fig:Zd_Rs}
\end{figure}

  At $r_s \geq r_s^{(c)}$ our numerical method diverges. To see what
  happens, we introduce an asphericity parameter
  \begin{equation}
    Z_{c} = \frac{ \int d^{4} x\, \sqrt{g_E} \; \varphi^4 \, z}
    {\int d^{4} x\, \sqrt{g_E} \;  \varphi^4}\,, \qquad z
    \equiv R \cos\theta
    \label{eq:Z_d}
  \end{equation}
  measuring coordinate distance between the BH origin and the
    center of the bubble interaction energy. The function~$Z_c (r_s)$
    is non-monotonic on the aspherical branch, see the solid and
    dashed red lines at ${r_s^{(a)} < r_s < r_s^{(c)}}$ in  
    Fig.~\ref{fig:Zd_Rs}. In particular,~${d Z_c/dr_s = -\infty}$
    at~$r = r_s^{(c)}$ indicates turn-around of a smooth
    curve: the inverse function~$r_s(Z_c)$ has a maximum~${dr_s/dZ_c =
      0}$ at this point  beyond which it should decrease.

We thus expect two types of aspherical bubbles at ${r_s <
    r_s^{(c)}}$  and none above this value. To find the second branch
  at~${r_s \approx r_s^{(c)}}$, we use the existing solutions as
  initial seeds. In this region we change parametrization: fix~$Z_c$
  in Eq.~(\ref{eq:Z_d}) to a value somewhat below~$Z_c(r_s^{(c)})$, but 
  adjust~$r_s$ when solving the equations. Details of this numerical
  procedure will be presented in Sec.~\ref{Sec:Lipatons}. Once it
  converges to a solution, we switch 
  back to the old method of stepping over~$r_s$ thus recovering the 
  entire second branch, see the dashed lines marked ``second
  aspherical'' in Figs.~\ref{fig:full_res} and~\ref{fig:Zd_Rs}.   

\begin{figure}[ht]
  \includegraphics{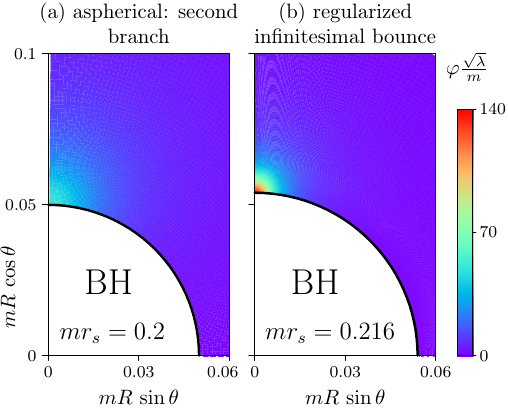}
  \caption{(a)~Aspherical critical bubble from the second
      branch at $r_s \equiv 4R_s =
      0.2\, m^{-1}$. (b)~Infinitesimal bounce at $r_s =
      0.216\, m^{-1}$ sitting on the event horizon
      and regulated by Eq.~\eqref{eq:V_reg} with~$\gamma_6 = 4 \cdot
      10^{-7}$.} 
  \label{fig:2nd_lip}
\end{figure}

Aspherical bubbles from the second branch are narrower and higher
  than from the first, cf.\ Figs.~\ref{fig:2nd_lip}(a)
  and~\ref{fig:aspher_0.05} bewaring the scales. Together, the two
  branches form a smooth family, see
  Fig.~\ref{fig:Zd_Rs}. But their suppressions in
  Fig.~\ref{fig:full_res} join at a cusp\footnote{This is a
    signal of extra negative mode emerging on the second
    branch. Indeed, let us parametrize the
    solutions~$\varphi_{\mathrm{cb}}^{\mathrm{(a)}}(\mbs{x})$ and~${r_s = 
      r_s(Z_c)}$ with~$Z_c$. Taking $Z_c$ derivative of the 
    equation for~$\varphi_{\mathrm{cb}}^{\mathrm{(a)}}$ at ${r_s = r_s^{(c)}}$
    and using ${dr_s /dZ_c = 0}$, we find out that the
    function~$\partial_{Z_c}\varphi_{\mathrm{cb}}^{\mathrm{(a)}} (\mbs{x})$ 
    is a  zero mode with~$\mu = 0$ in the background of this solution,
    cf.\ Eqs.~\eqref{eq:sph} 
    and~\eqref{eq:eigen_combo}. Thus, one of the mode
    eigenvalues crosses zero at~${r_s = r_s^{(c)}}$.}~${r_s 
    = r_s^{(c)}}$ with~${dS_E / dr_s \ne 0}$ and hence~${dS_E / dZ_c 
    = dr_s / dZ_c = 0}$. It is also important that solutions from the
  second branch have  larger suppressions~$S_E$  and hence
  subdominant.  

  To sum up, we have found a complete smooth family of aspherical
  bubbles, identified their dominant branch and argued that solutions 
  from the latter branch have only one negative mode. Unfortunately,
  the new bubbles exist only at $r_s < r_s^{(c)}$. It is not clear how to 
  describe transitions near larger black holes, something is missing.
	
%%%%%%%%%%%%%%%%%%%%%%%%%%%%%%%%%%%%%%%%%%%%%%%%%%
        
\section{Infinitesimal bounces in Schwarzschild spacetime}
\label{Sec:Lipatons}	

  Now, we explore families of static bubbles which are completely
  disconnected from the ones we have gotten so far. We
  compute them using the following trick. At the first stage, we
  extremize~$S_E$ in the class of configurations with given
  asphericities~$Z_c$. This is equivalent to computing saddle
  points  of the functional
  \begin{gather}
    \label{eq:S-lC}
    \mathcal{F}[\vf] = S_E[\vf] - \mu_{Z} \int d^4 x \sqrt{g_E} \,
    (z - Z_c) \varphi^4\,,
  \end{gather}
  where extremization $\partial {\cal F}/\partial \mu_Z =
  0$ with respect to the Lagrange multiplier~$\mu_Z$ imposes
  the constraint~(\ref{eq:Z_d}). The outcome of the first stage is a
  smooth family of saddle-point configurations $\varphi_{Z}(\mbs{x})$
  parametrized with~$r_s$ and~$Z_c$. At the second stage we manually
  extremize $S_E[\varphi_Z]$  with respect to~$Z_c$ and thus get critical 
  bubbles $\varphi_{\mathrm{cb}}(\mbs{x})$ satisfying the original
  field equation. They may be disconnected from other~$S_E$ saddles. 

\begin{figure}
  \includegraphics{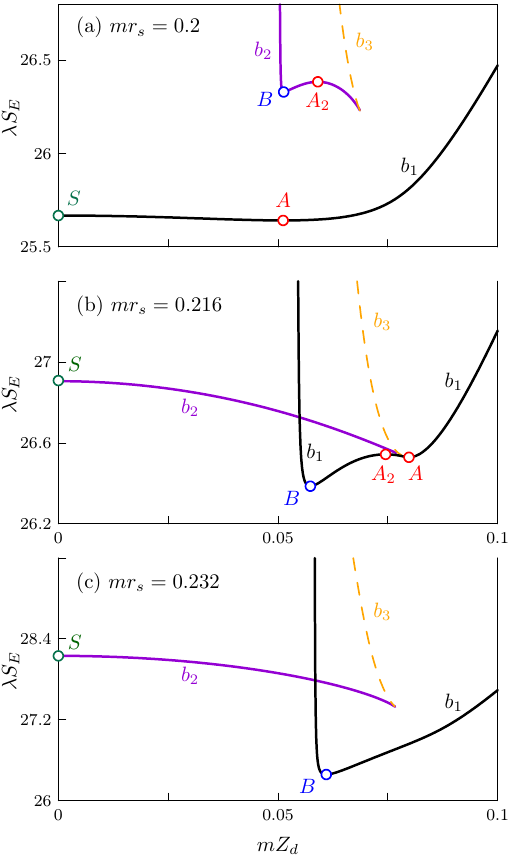}
  \caption{Suppressions $S_E[\varphi_Z]$ of solutions with
      fixed~$Z_c$ at two values of~$r_s$ shown by arrows in
      Fig.~\ref{fig:Zd_Rs}: (a)~${r_s = 0.2\, m^{-1}}$,
      (b)~${r_s = 0.216\, m^{-1}}$, and (c)~${r_s = 0.232\, m^{-1}}$. We use
      regularization~\eqref{eq:V_reg}.}
  \label{fig:branches}
\end{figure}

  Before diving into numerical analysis, we add tiny
  regulator to the potential~\eqref{eq:pot}: 
  \begin{equation}
    V(\varphi) \to V(\varphi) + \lambda_6 \, \frac{\vf^6}{6}\,,
    \qquad \lambda_6 = \gamma_6 \; \frac{\lambda^2}{m^2} \,, 
    \label{eq:V_reg}
  \end{equation}
  where $\gamma_6 = 4 \cdot 10^{-7} > 0$. It smoothes singular
  solutions with infinite~$\varphi$ but leaves the others almost intact. 

  Figure~\ref{fig:branches} shows implementation of~$Z_c$
  fixation procedure. We launch it from every critical 
  bubble we know: spherical~$S$, aspherical~$A$, and second 
  aspherical~$A_2$. Changing~$Z_c$ in small steps, we repeatedly solve
  saddle-point equations~$\delta {\cal F}/\delta \varphi =
  \partial {\cal F}/\partial \mu_Z = 0$ using previous solution as 
  initial approximation for the next. Computation ends when all
  available solutions are found. 
	
In this way we discover two continuous families of static
  configurations~$\varphi_Z (\mbs{x})$ at every~$r_s$. Their structure
  changes with BH radius. At small~$r_s$ the bubbles~$S$
  and~$A$ sit on a single branch while~$A_2$ belongs to another,
  completely subdominant one, see Fig.~\ref{fig:branches}(a). The
  suppression~$S_E$ of the dominant branch, however, grows with~$r_s$
  until it reaches the subdominant contribution. Eventually, the two
  branches reconnect, see Fig.~\ref{fig:branches}(b). At $r_s \approx
  r_s^{(c)}$ the dominant branch (solid line)  
  includes aspherical bubbles~$A$ and~$A_2$, whereas the subdominant
  one (dashed) accommodates~$S$. At~$r_s > r_s^{(c)}$ the extrema~$A$
    and~$A_2$ vanish, but in other respects the branches retain their forms, as seen in Fig.~\ref{fig:branches}(c).

  It is worth noting that the second branch of aspherical critical
  bubbles was obtained in Sec.~\ref{Sec:Aspherical} using
  Fig.~\ref{fig:branches}(b) with~${r_s \approx r_s^{(c)}}$. We 
  computed the solution~$A_2$ starting from~$A$ and stepping
  over~$Z_c$. Then we relaxed the constraint and continued 
  changing~$r_s$.  

  Note also that the limit~$Z_c\to +\infty$ of constrained solutions
  yields flat-space critical bubbles far away from the black
  hole. They have higher energies and suppressions due to vanishing
  mass defect in the  black hole field, see the rightmost part of 
  Fig.~\ref{fig:branches}.

  We are interested in extrema of~$S_E[\varphi_Z]$ with respect
  to~$Z_c$, since they represent static Euclidean solutions, critical
  bubbles. There are four of them in Fig.~\ref{fig:branches}: $S$, 
  $A$, $A_2$, and~$B$,  plus a cusp in the right-hand side of the
  plot. Within this set, the second aspherical bubble~$A_2$
    and the cusp are always subdominant and hence uninteresting. The other 
  extrema are dominant with the smallest~$S_E$ at certain values of~$r_s$, see
  Eqs.~\eqref{eq:intervals_rs}. Namely, as the BH radius grows from
  zero, spherically-symmetric bubble~$S$ dominates at first,  then it
  gives this role to the aspherical solution~$A$, until the new 
  saddle point~$B$ starts to dominate at large~$r_s$. This behavior is
  responsible for piecewise-smooth suppression exponent in
  Fig.~\ref{fig:intro_answer} and three mechanisms of
  false vacuum decay.

  Now, to the best part: let us disclose the nature of mysterious static
  bubble~$B$. Amazingly, this is a Fubini-Lipatov
  instanton~\cite{Fubini:1976jm, Lipatov:1976ny, Affleck:1980mp,
    Linde:1981zj}, a bounce of negligibly small size~$a$ reviewed in
  Sec.~\ref{Sec:Flat}. Since the bounce lives in four dimensions and
  has $O(4)$ symmetry, it
  is not immediately clear how it squeezed itself into the ranks of
  static aspherical bubbles. 

\begin{figure}[ht]
  \includegraphics{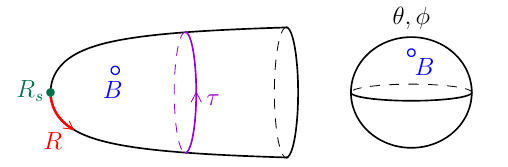}
  \caption{Euclidean Schwarzschild spacetime (not to scale).  
    Filled dot shows the event horizon $R = R_s$, empty
    circle is the bounce~$B$ sitting at an arbitrary
      spacetime point.} 
  \label{fig:menzoorka}
\end{figure}

  Recall, however, that Euclidean black hole is a smooth manifold
  looking like a cigar for~$R$ and~$\tau$ coordinates equipped with
   $(\theta,\, \phi)$ spheres, see Fig.~\ref{fig:menzoorka}. It is 
  locally flat everywhere including the horizon at~${r_s \equiv
    4R_s}$. Thus, the flat-space bounce~\eqref{eq:liprofile} with~$a
  \to 0$ obeys field 
  equation near an arbitrary point of this manifold if it is written in
  locally flat coordinates. Indeed, this configuration is localized and
  therefore feels spacetime curvature as power corrections in~${a^2/R_s^2
    \to 0}$. The same logic guarantees that the bounce
  action~\eqref{eq:flat_action} is equal to its flat-space value $S_b\equiv
  8\pi^2/(3\lambda)$ in  Eq.~\eqref{eq:lipaton} irrespectively of the
  bounce location. Finally, the bounce field~\eqref{eq:liprofile} has~$O(4)$ 
  symmetry and perceived as static ($\tau$-independent) if located
  precisely on the horizon $R = R_s$.

  Let us demonstrate the above properties explicitly. Suppose the
  bounce sits at the point~${R \approx R_s}$ and~$\theta \approx 0$ of
  the horizon. We
  introduce locally flat coordinates near this point,
\begin{equation}
  \widetilde{\tau} = \tau/8R_s\,,\quad 
    \varrho \approx  4(R - R_s)\,, \quad
    \varrho_\theta = 4R_s \theta\,,
    \label{eq:rho_horizon}
  \end{equation} 
    where~$\varrho$ and $\varrho_{\theta}$ are small and
    $\widetilde{\tau}$ changes by~$2\pi$ every $\beta$-period, 
    see Eqs.~(\ref{eq:T_H}), (\ref{eq:isotropic_r}). The
    metric~(\ref{eq:isotropic_metric}) near the horizon takes the form 
    \begin{equation}
      ds_E^2 \approx \left(\varrho^2 d{\widetilde{\tau}}^2 + d\varrho^2\right)
      + \left(d\varrho_\theta^2 + \varrho_\theta^2d \phi^2 \right)\,,
		\label{eq:metric_horizon}
    \end{equation}
    where corrections in~$\varrho, \varrho_\theta \ll R_s$ are
    omitted. It indeed describes flat space $\mathbb{R}^2\times
    \mathbb{R}^2$ in double-polar coordinates $(\varrho, 
    \widetilde{\tau})$ and $(\varrho_\theta,\, \phi)$. 
	
  Renotating the coordinates, we write Fubini-Lipatov
  instanton~\eqref{eq:liprofile} as
  \begin{equation}
    \varphi_b\approx \sqrt{\frac{8}{\lambda}}\,
    \frac{a}{a^2+\varrho^2 + \varrho_\theta^2}\quad \mbox{at}\quad  a \ll R_s
    \mbox{ and } m^{-1}\,,
    \label{eq:liprofile2}
  \end{equation}
  where in truth,~$a\to 0$ in the model we consider, see
  Eq.~\eqref{eq:m-bounce}. Now, one can explicitly check that
  Eq.~(\ref{eq:liprofile2}) satisfies field equation up to
  corrections in~$a/R_s$, see calculations in
  Appendix~\ref{App:ReguLip}. It is also obvious  that the solution
  with~${a=0}$ has flat-space Euclidean action~${S_E[\varphi_b] =
    S_b}$. Remarkably, $\varphi_b$ in Eq.~\eqref{eq:liprofile2} is
  time-independent and  axially-symmetric, yet aspherical, as it
  depends on~${\theta \propto 
    \varrho_\theta}$. These are the properties of the numerical
  extremum~$B$ obtained in Fig.~\ref{fig:branches}.

  Now, we prove that the new critical bubble~$B$ is the
  bounce~\eqref{eq:liprofile2}. To this   
  end we recall that our numerical calculus was performed in the
  regularized model~(\ref{eq:V_reg}) with no singular saddle-point
  solutions. Regularization stabilizes the bounce size at 
  \begin{equation}
    a \propto  \gamma^{1/4}_6 / m\,,
    \label{eq:a_g6}
  \end{equation}
  and adds correction to its classical action,
  \begin{equation}
    \delta_S \equiv S_b^{\mathrm{BH}}/S_b - 1
    \approx  d_b \sqrt{\gamma_6}  + O(\gamma_6)\,,   
    \label{eq:a_R_4}
  \end{equation}
  making it sensitive to the metric, see derivation in
  Appendix~\ref{App:ReguLip} and cf.\ Ref.~\cite{Affleck:1980mp}.

\begin{figure}[ht]\centering
	\unitlength=1mm
	\begin{picture}(86,61)
		\put(0,0){\includegraphics{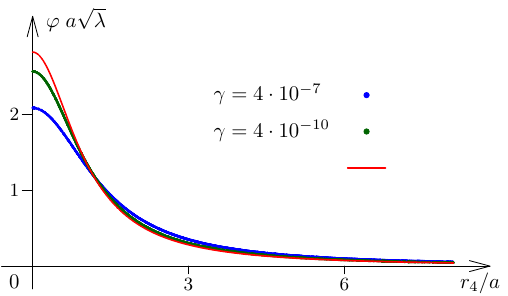}}
		\put(44,20.7){Eq.~\eqref{eq:liprofile2}}
	\end{picture}
  \caption{The field~$\varphi_{\mathrm{cb}}(R,\theta)$ of
      near-horizon bounce in Fig.~\ref{fig:2nd_lip}(b) versus the 
      geodesic distance~$r_4 \equiv \sqrt{\varrho^2 +
        \varrho_\theta^2}$ to the horizon point~$R = R_s$ and~$\theta=0$;
      see Eq.~(\ref{eq:rho_horizon}). Numerical results (dots) are
      compared to Eq.~(\ref{eq:liprofile2}) (line). We use $r_s = 0.216
      \, m^{-1}$, compare two different regulators~$\gamma_6 = 4 \cdot 10^{-7}$ and $\gamma_6 = 4 \cdot 10^{-10}$, and
      extract~$a\propto \gamma^{1/4}/m$ from
      analytic equation~(\ref{eq:a_g6_full}) of Appendix~\ref{App:ReguLip}.}
  \label{fig:bounce_profile}
\end{figure}

\begin{figure}[ht]
  \includegraphics{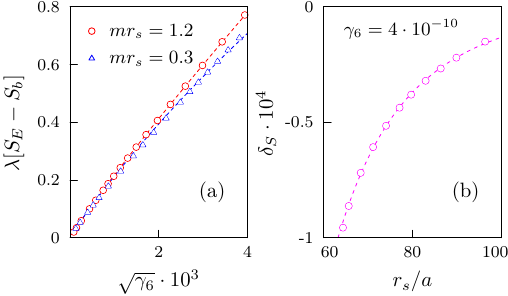}
  \caption{(a)~Euclidean action $S_b^{\mathrm{BH}}$ of the
      bounce~$B$ in Fig.~\ref{fig:branches} as a function of  the
      regulator~$\gamma_6$ at~$r_s = 1.2 \, m^{-1}$ (points) and $r_s = 0.3 m^{-1}$ (triangles). Lines
      show the theory~\eqref{eq:a_R_4} with~$d_b$ given by
      Eq.~\eqref{eq:d_b_W} and~$O(\gamma_6)$ contribution extracted from the fit. (b)~Relative   
      difference $\delta_S \equiv
      S_b^{\mathrm{BH}}/S^{\mathrm{flat}}_b - 1$ of two Euclidean
      actions of the bounces. One of them is~$B$ sitting on the BH
      horizon and another one is computed in flat
      spacetime;~${\gamma_6  = 4 \cdot 10^{-10}}$. Dots represent numerical data, line depicts the $(a/r_s)^4$ fit.  }
  \label{fig:horiz_vs_flat}
\end{figure}

  The bubble~$B$  is displayed at~${\gamma_6 = 4 \cdot 10^{-7}}$ in
  Fig.~\ref{fig:2nd_lip}(b). It is indeed small in size, squeezes to a
  point as~$\gamma_6 \to 0$, and its profile in 
  Fig.~\ref{fig:bounce_profile} (dots) is close to 
  Eq.~(\ref{eq:liprofile2}) (line), getting even closer to it as the regulator decreases, e.g., to ${\gamma_6 = 4 \cdot 10^{-10}}$. Moreover, the Euclidean action of this bubble
  approaches~$S_b$ as~$\gamma_6$ decreases, see
  Fig.~\ref{fig:horiz_vs_flat}(a). We conclude that~$B$ is
  the Fubini-Lipatov instanton (\ref{eq:liprofile2}) itself. In the
  unregularized theory with   
  $\gamma_6 = 0$ it is infinitely thin and tall, sits at the horizon
  point with $R = R_s$, and is suppressed by the action $S_b \equiv   
  8\pi^2/3\lambda$. This solution is  dominant at $r_s >  r_s^{(b)}$,
  see Eq.~(\ref{eq:lip_dom}) and Fig.~\ref{fig:intro_answer}. 

  To finish we remark that bounces in our original model
  with~${\gamma_6 = 0}$ are atypical. Being singular, they have the
  same suppression~$S_b$ irrespective of the position in curved
  Euclidean spacetime. We expect that in models with bounded
  potentials this degeneracy will be lifted leaving only two finite-size
  solutions: bounce sitting on the BH horizon\footnote{Existing in
    certain region of~$r_s$.  In our regularized model with~$\gamma_6
    > 0$ the branch of horizon-riding bounces terminates at small $r_s$ by forming a cusp with the subdominant
    branch of aspherical bubbles.\label{fn:2nd_lip_bifur}} $R = R_s$ and another one living in flat space at~$R\gg
  R_s$. Only one of them will be dominant. For 
  example, in our model with finite~$\gamma_6 > 0$ the near-horizon
  bounce always has smaller suppression~$S_E[\varphi_{\mathrm{cb}}]$,
  see Fig.~\ref{fig:horiz_vs_flat}(b). This is natural: after all,
  the black hole attraction lowers potential barrier for tunneling. We
  therefore anticipate that vacuum tunneling generically happens near
  event horizons.

%%%%%%%%%%%%%%%%%%%%%%%%%%%%%%%%%%%%%%%%%%%%%%%%%

\section{Discussion}
\label{Sec:Discussion}
In this paper we studied decay of a false scalar field vacuum 
  near a black hole equilibrated at Hawking temperature $T = (4\pi
  r_s)^{-1}$ with an environment. We gave general argument
  that this process is rotationally asymmetric with respect to the black
  hole center unless the latter is very small and hot:~$r_s > r_s^{(a)}
  \propto m^{-1}$, where~$r_s$ is the black hole radius and~$m$ is the
  field mass. 

  We confirmed this asphericity  in the model~\eqref{eq:pot} by
  computing dominant semiclassical solutions and exponential
  suppression~$S_E \sim \ln
  \Gamma$ of the decay rate~(\ref{eq:exp-S}). At $r_s<
  r_{s}^{(b)}$ in Eq.~(\ref{eq:intermediate_bh}) our solutions
  describe activation, that is formation of unstable critical bubbles with
  embryos of true vacuum 
  inside. Specifically, at $r_s<  r_{s}^{(a)}$ in
  Eq.~(\ref{eq:small_bh}) the bubbles are
  spherical and cover the entire horizon, while 
  at $r_s^{(a)} < r_s < r_s^{(b)}$ they become small, rotationally
  asymmetric and squeezed to the black hole side, indeed; see
  Fig.~\ref{fig:decay_schematic}. This is 
  an explicit example of dominant semiclassical solutions breaking 
  $O(3)$ symmetry of the background, cf.\ Ref.~\cite{Avraham:2026mgk}. 

  At $r_s > r_s^{(b)}$ (low temperature) the decay mechanism changes
  to vacuum tunneling via Euclidean
  bounces~\cite{Coleman:1977py, Coleman:1978ae}. In our
  model~(\ref{eq:pot}) the bounces are very
  unusual~\cite{Fubini:1976jm, Lipatov:1976ny, Affleck:1980mp, 
    Linde:1981zj}: have infinitesimally small sizes and
  suppression~$S_E = 8\pi^2/(3\lambda)$ insensitive to their locations
  in curved spacetime. We demonstrated that in  regularized
  model~(\ref{eq:V_reg}) with finite-size bounces this degeneracy is
  lifted and the rotational symmetry is broken: vacuum tunneling is
  aspherical and happens  near the point of the event horizon.

  Suppression exponent~$S_E$ of the resulting decay rate is shown in
  Fig.~\ref{fig:full_res} as a function of the black hole radius~$r_s$
  (solid lines). It is piecewise smooth because the transition
  mechanisms switch from spherical activation to aspherical
  activation and then to vacuum tunneling as~$r_s \propto T^{-1}$
  grows above~$r_s^{(a)}$ and~$r_s^{(b)}$. The latter two critical
  radii correspond to Hawking temperatures 
  \begin{equation}
    T^{(a)}_{\mathrm{cr}} \approx 0.41 \, m\quad \mbox{and} \quad
    T^{(b)}_{\mathrm{cr}}\approx 0.377 \, m\,,
  \end{equation}
  where Eq.~(\ref{eq:T_H}) is used.

  Our results suggest that black hole-induced false vacuum decay may
  be aspherical in other setups and models. This calls for
  reexamination of decay rates near black holes equilibrated with
  thermal baths~\cite{Tetradis:2016vqb, Briaud:2022few}, black
  holes immersed in baths with non-matching
  temperatures~\cite{Hu:2026udq, 
    Gazizov:2026}, and isolated black holes~\cite{Gregory:2013hja,
    Burda:2016mou, Gorbunov:2017fhq, Mukaida:2017bgd, Kohri:2017ybt,
    Hayashi:2020ocn, Shkerin:2021zbf, Shkerin:2021rhy,
    Strumia:2022jil, Gazizov:2026}. 

  The topnotch application of our results is finite-temperature
  decay of false Higgs vacuum near black holes. At large fields, the
  Higgs potential looks just like our Eq.~\eqref{eq:pot}, but with
  essential differences. Its mass term with~$m = m_H$ is negligible as 
  compared to self-interaction energy density of the fields $\varphi \sim 10^{9} \,
  \opn{GeV}$ emerging during the decay. Nevertheless,  the scale invariance of
  the model is broken by the running four-coupling~$\lambda =
  \lambda_H(\varphi)$~\cite{Bednyakov:2025uur}. In addition, thermal  
  corrections~\cite{Anderson:1990aa, Arnold:1991cv, DelleRose:2015bpo,
  Strumia:2022jil} drastically modify the  potential at high
  temperatures. All these effects may alter intricate structure of 
  aspherical semiclassical contributions which deserve separate
  careful study.

  One can advance in two other, more technical directions. First, our
  arguments that the aspherical bubbles and bounces are physical with
  only one negative mode were indirect. It would be supportive to
  compute the eigenspectra of the solutions and demonstrate this
  explicitly. Second, the problem of including (weak)
  backreaction of the 
  aspherical scalar 
  field on the geometry is a fascinating task for
  passionate experts,
  cf.\ Refs.~\cite{Gregory:2013hja, Burda:2016mou, Mukaida:2017bgd}.

\begin{acknowledgments}
  This study was conducted within the framework of a scientific
  program of the National Center for Physics and Mathematics,
  section~5 ``Particle Physics and Cosmology,'' stage
  2026--2027. Numerical calculations were performed on the
  Computational Cluster of Theoretical Division of INR RAS. 
\end{acknowledgments}

	\appendix

%%%%%%%%%%%%%%%%%%%%%%%%%%%%%%%%%%%%%%%%%%%%%%%%

\section{Small-size bounces}
\label{App:ReguLip}

In this Appendix we study bounces in the original~\eqref{eq:pot}
  and regularized~\eqref{eq:V_reg} models. These objects have small
  cores of size~$a\ll m^{-1}, r_s$ and long tails. We describe them using
  matched asymptotic expansions in the two regions and extremization
  of~$S_E$ with respect to free parameters. 

%%%%%%%%%%%%%%%%%%%%%%%%%%%%%%%%%%%%%%%%%%%%%%%%

\subsection{Flat-space Fubini-Lipatov instantons}
\label{sec:original-model-flat}
	
Start with the model~\eqref{eq:pot} in flat space. The respective 
   bounce~$\varphi_b$ is~$O(4)$ symmetric, i.e.\ depends only on
  Euclidean four-radius $r_4 = \sqrt{\tau^2 + \mbs{x}^2}$.  We 
  introduce its overlapping core and tail at~${r_4  \ll m^{-1}}$
  and~${r_4 \gg a}$. Inside the core, the mass term in
  Eqs.~(\ref{eq:flat_action}), (\ref{eq:pot}) is suppressed. Indeed,
  rescaling with the bounce size~$a$, 
  \begin{equation}
    \label{eq:4}
    \varphi_b = \frac{\sqrt{8}}{a\sqrt{\lambda}}\; \varphi'_b (r_4') \,, \qquad 
    r_4 = a r'_4\,,
  \end{equation}
  brings the field equation to the form
  \begin{equation}
    \label{eq:5}
    \prt_{r'_4}^2 \varphi'_b + \frac{3}{r'_4} \,\prt_{r'_4} \varphi'_b +
    8\varphi'^3_b =  (ma)^2 \varphi'_b
  \end{equation}
  exposing~${(ma)^2 \ll 1}$ as a
  natural expansion parameter:  
  \begin{equation}
    \varphi'_b = \varphi'_0(r'_4)  + (ma)^2 \varphi'_1(r'_4) +  O\left(ma\right)^4 \,.
    \label{eq:inner_expansion}
  \end{equation}
  Here $\varphi_0$ is the Fubini-Lipatov instanton~\eqref{eq:liprofile},
  \begin{equation}
    \label{eq:FL_rescaled} 
    \varphi'_0(r'_4) = (1+r'^2_4)^{-1}\,,
  \end{equation}
  and $\varphi_1$ satisfies equation
  \begin{equation*}
  \left(\prt_{r'_4}^2 + \frac{3}{r'_4} \,\prt_{r'_4} +
    24 {\varphi_0'}^2 \right) \varphi'_1 = \varphi'_0\,.
  \end{equation*}
  with no analytic solution. However, the asymptotics
\begin{equation}
  \varphi'_1 = \frac{1}{2} \ln r'_4 + \frac{c_0'}{2} + O(r_4')^{-2}
  \quad \mbox{at} \quad r_4' \gg 1\,,
  \label{eq:psi_1_asymp}
\end{equation}
is fixed by the equation up to an unknown constant~$c_0'$. 

  In the tail region~${r_4 \gg a}$ the field mass is important, but the
  amplitude is already small. Changing the variables,
  \begin{equation}
    \label{eq:2}
    \varphi_b = \frac{am^2}{\sqrt{\lambda}} \;
    \tilde{\varphi}_b(\tilde{r}_4)\,, \qquad 
    r_4 = \tilde{r}_4/m\,,
  \end{equation}
  we indeed see that the small parameter~${(ma)^2 \ll 1}$ appears in
  front of the cubic term in the equation,
  \begin{equation}
    \label{eq:1}
    \prt_{\tilde{r}_4}^2\tilde{\varphi}_b  + \frac{3}{\tilde{r}_4} 
      \prt_{\tilde{r}_4} \tilde{\varphi}_b - \tilde{\varphi}_b  =  -(ma)^2
    \tilde{\varphi}^3_b \,,
  \end{equation}
  see Eqs.~(\ref{eq:flat_action}), (\ref{eq:pot}). Imposing vanishing
  boundary conditions at infinity, we obtain the leading-order
  solution
  \begin{equation}
    \label{eq:K1_solution}
    \tilde{\varphi}_b = \tilde{c}_0 K_1(\tilde{r}_4) / \tilde{r}_4  + O(ma)^2\,,
  \end{equation}
  where $\tilde{c}_0$ is an arbitrary constant and $K_1$ is the modified
  Bessel function. 

  Next, we match the two solutions in the overlap region $a
  \gg r_4 \gg m^{-1}$ with large~$r_4' $ and small~$\tilde{r}_4$.
  To this end we compare~${\tilde{r}_4 \to 0}$ asymptotics
  of the tail~(\ref{eq:K1_solution}),
  \begin{equation}
    \tilde{\varphi}_b = \frac{\tilde{c}_0}{2}
    \left[\frac{2}{\tilde{r}^2_4} + \ln \frac{\tilde{r}_4}{2} +
      \gamma_E - \frac12 \right] + O\left(\tilde{r}_4\right)^2\,,
    \label{eq:chi_0_asymp}
  \end{equation}
  with $r_4' \to \infty$ limit of the core
  solution~\eqref{eq:inner_expansion}, 
  \eqref{eq:FL_rescaled}, \eqref{eq:psi_1_asymp} paying attention to 
  the rescalings~\eqref{eq:4} and~(\ref{eq:2});~${\gamma_E \approx
    0.577}$ is the Euler's constant. This specifies the coefficients
  \begin{equation}
    \label{eq:3}
  c_0' = \ln (ma/2) + \gamma_E - \frac12  \quad\mbox{and} \quad
  \tilde{c}_0 = \sqrt{8}\,.
\end{equation}
  We arrived at the leading-order bounce
  solution~(\ref{eq:FL_rescaled}), (\ref{eq:psi_1_asymp}),
  (\ref{eq:K1_solution}), and~(\ref{eq:3}) in the framework of
  two-region expansion. If needed, this procedure can be continued to
  find higher-order corrections in~$(ma)^2$.

  The above expansion has a free parameter: the bounce size~$a$. We
  fix it by evaluating the bounce action~$S_E[\varphi_b]$~--- a sum of
  contributions from the core~${r_4< r_*}$ and tail~$r_4>r_*$ regions,
  where 
  $$
  a \ll r_* \ll m^{-1}\,.
  $$
  First, the core. Performing rescaling~\eqref{eq:4} and  expansion~(\ref{eq:inner_expansion}) in the integrand of flat-space
  Eq.~(\ref{eq:flat_action}), we~get, 
  \begin{align}
    \notag
    \frac{\lambda S^{\mathrm{core}}_E}{8 \pi^2}  = &
    \int\limits_0^{r_*/a} r'^3_4 dr'_4  \left[(\prt_{r'_4}
    \varphi'_0)^2 - 4\varphi'^4_0 + (ma)^2
    \varphi'^2_0\right] \\ \notag
    & \quad + 2 (ma)^2 \; r'^3_4 \varphi'_1 \prt_{r'_4}
    \varphi'_0\Big|^{r_*/a}_0 + O(ma)^4\,, \\\notag
    = & \frac13 - \frac{2a^2}{r_*^2} -  (ma)^2 \ln 
    \frac{ar_* m^2}{4}
    +   (ma)^2 \frac{1  - 4 \gamma_E}{2} \\ & +
    O(a^4) + O(r_*)^{-4}\,,
    \label{eq:S_M_core}
  \end{align}
  where the terms with $\varphi_1'$ were integrated by parts, we used
  Eqs.~\eqref{eq:5}, ~(\ref{eq:FL_rescaled}), and (\ref{eq:psi_1_asymp})
  and packed corrections proportional to~$a^4$ into~$O(a^4)$. 

  Second, contribution from the tail region is given by quadratic part of
  the action~(\ref{eq:flat_action}) at $r>r_*$, 
  \begin{align}
    \label{eq:6}
    \frac{\lambda S^{\mathrm{tail}}_E}{8\pi^2} = &\;   \frac{(ma)^2}{8}\;
    \tilde{r}^3_4\,  \tilde{\varphi}_b \,\prt_{\tilde{r}_4} \tilde{\varphi}_b
    \Big|_{mr_*}^{\infty}  + O(a^4)\\ \notag
    = & \; \frac{2a^2}{r_*^2} \!+
    (ma)^2 \left[\ln \frac{mr_*}{2} +\gamma_E -
      1\right] + O(a^4)\,,
  \end{align}
  where we again integrated by parts and exploited Eqs.~\eqref{eq:1}, 
  \eqref{eq:chi_0_asymp}.

  Adding up Eqs.~\eqref{eq:6} and~(\ref{eq:S_M_core}), we obtain 
  bounce action~\eqref{eq:m-bounce} with parameter
  \begin{equation}
    \frac{c_b}{3} = \ln \frac{2}{ma} - \frac12 - \gamma_E > 0
    \label{eq:mass_corr_answer}
  \end{equation}
  which is logarithmically sensitive to~$ma$. Note that dependence
  on the arbitrary scale~$r_*$ canceled, as it should. 

  The size~$a$ of the dominant bounce  is discussed in the main
  text. In our unregularized model minimal suppression~$S_E[\varphi_b]$ is
  achieved in the limit~$a\to 0$ which  corresponds to infinitezimally
  thin and infinitely high bounce. This gives the above solutions the
  sense of constrained instantons~\cite{Affleck:1980mp}.  

%%%%%%%%%%%%%%%%%%%%%%%%%%%%%%%%%%%%%%%%%%%%%%

\subsection{Regularized bounces in flat space}
\label{sec:regul-model-flat}

  To stabilize  bounce size, we added the regulator~\eqref{eq:V_reg} 
  with~${\gamma_6 \ll 1}$ to the scalar potential. The respective change of
  the flat-space bounce action can be obtained by substituting
  the unregularized solution from
  Appendix~\ref{sec:original-model-flat} into the regulator action, 
  \begin{equation} 
    \frac{\Delta S_E^{\gamma_6}}{S_b}\equiv \frac{\gamma_6
      \lambda^2}{6m^2 S_b}  \int d^4 x
    \, \varphi_b^6  = 
    \frac{8\gamma_6}{5(ma)^2}  + O(\gamma_6)\,,
    \label{eq:g6_corr}
  \end{equation}
  where we used Eqs.~(\ref{eq:lipaton}) and~(\ref{eq:FL_rescaled})
  ignoring~$\gamma_6(ma)^0$  
  corrections. Indeed, $\gamma_6$~shift of the bounce profile 
  adds negligible~$\gamma_6^2$ terms to~$S_E$: the original action
  is extremal with respect to~$\varphi_b$ whereas
  the new term~\eqref{eq:g6_corr} is proportional to~$\gamma_6$ itself.  This
  means that the full action of the regularized bounce is the sum of 
  Eqs.~\eqref{eq:m-bounce} and~\eqref{eq:g6_corr}, 
  \begin{equation}
    \label{eq:SE_a_g6}
    \frac{S_E[\varphi_b]}{S_b} = 1 + c_b(a) (ma)^2  +
    \frac{8\gamma_6}{5(ma)^2} + O(ma)^4  .
  \end{equation}
  It is minimal if $a$ satisfies equation
  \begin{equation}
    (ma)^4 \left[\ln \frac{2}{ma} -1 - \gamma_E \right] = \frac{8}{15}
    \gamma_6\,,
    \label{eq:a_g6_full}
  \end{equation}
  where we used Eq.~(\ref{eq:mass_corr_answer}). The respective
  saddle-point action has the form~\eqref{eq:a_R_4}
  with
  \begin{equation} 
    \label{eq:d_b_for_action}
    d_b  = \frac{8 c_b - 6}{\sqrt{10c_b - 15}}\,.
  \end{equation}
  Note that~$c_b$, $d_b$, and~$a$ are\footnote{In Fig.~\eqref{fig:bounce_profile} we
  		use numerical solution of Eq.~(\ref{eq:a_g6_full}).} logarithmically sensitive
  to~$\gamma_6$ via the Lambert $W$-function:
  \begin{align}
  	a = \frac{1}m \left[ \frac{-\frac{32}{15}\, \gamma_6}{W_{-1}\left(\varpi \gamma_6 \right)}\right]^{1/4} 
  	\label{eq:a_W}
  	, \\
  	d_b = \frac32 \left[ \frac{-\frac{32}{15}\, \gamma_6}{W_{-1}\left(\varpi \gamma_6 \right)}\right]^{1/2} \left(1 - W_{-1}\left(\varpi \gamma_6 \right)\right),
  	\label{eq:d_b_W}
  \end{align}
  where $\varpi \equiv e^{4(1+\gamma_E)}$ and for small negative arguments $W_{-1}(x) \simeq \ln (-x) - \ln (-\ln(-x))$. 
  Thus we reproduced Eqs.~\eqref{eq:a_g6} and~\eqref{eq:a_R_4} from the main
  text with coefficients including $\ln\gamma_6$.

\begin{figure}
  \centering
  \unitlength=1mm
  \begin{picture}(80,55)
    \put(0,0){\includegraphics{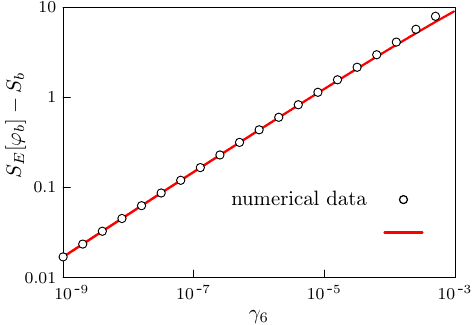}}
    \put(62.5,15.5){\makebox(0,0)[r]{Eqs.~\eqref{eq:a_R_4},
        \eqref{eq:a_g6_full}, \eqref{eq:d_b_for_action}}}
  \end{picture}
  \caption{Change~$S_E[\varphi_b] - S_b$ of the flat-space bounce
      action due to regularization~\eqref{eq:V_reg}.
    Solid line and circles show theoretical
    prediction~\eqref{eq:a_R_4}, \eqref{eq:a_g6_full}, 
    \eqref{eq:d_b_for_action} and numerical results, respectively.}
  \label{fig:S_m_g6}
\end{figure}

  In Fig.~\ref{fig:S_m_g6} we compare theoretical
  prediction~\eqref{eq:a_R_4}, \eqref{eq:a_g6_full},
  \eqref{eq:d_b_for_action} with numerically computed
  flat-space bounce action in the regularized 
  model~\eqref{eq:V_reg}, see Appendix~\ref{App:Modes} for
    details. The two results coincide at small~${\gamma_6}$, as they
  should.

%%%%%%%%%%%%%%%%%%%%%%%%%%%%%%%%%%%%%%%%%%%%%%%%

\subsection{Regularized bounces near large black holes}
\label{sec:regul-bounc-schw}

We proceed by placing the small-size bounce onto the black horizon in
  the regularized model~(\ref{eq:V_reg}), see
  Fig.~\ref{fig:menzoorka}. We will assume that the black hole is 
  large,~${r_s  \gg m^{-1}}$, and discuss generalization to
  smaller~$r_s$ afterwards. This approach is simple because
  well-localized bounces barely feel the spacetime curvature. 

  Indeed, let us introduce Riemann normal coordinates~$y^\mu$ in the
  vicinity of the bounce~\cite{Misner:1973prb}. They make the
  metric~(\ref{eq:isotropic_metric}) locally flat,
  \begin{equation}
    \label{eq:7}
    g_{E\, \mu\nu}(y)  =   \delta_{\mu\nu} - \frac{1}{3} {\cal
      R}_{\mu\lambda\nu\rho}^{(0)} \; y^\lambda y^{\rho}  + O(y^4)\,, 
  \end{equation}
  where~${y=0}$ is the horizon point,~${r_4  \equiv \sqrt{y_\mu
      y^{\mu}}}$ measures geodesic distance to it, while~${\cal
    R}^{(0)}_{\mu\nu\lambda\rho} \propto r_{s}^{-2}$ is a 
  Riemann tensor at $y=0$. Corrections to Eq.~\eqref{eq:7} are of
  order~$(y/r_s)^4$ due to parity symmetry~${y\to -y}$ of
  the Schwarzschild spacetime, see Fig.~\ref{fig:menzoorka} again. 

  Expansion~(\ref{eq:7}) suggests that  the shift of the bounce action due
  to gravity is proportional to~${{\cal R}^{(0)} \propto
    r_s^{-2}}$. It can be evaluated via the same strategy as in
  Sec.~\ref{sec:regul-model-flat}: by plugging  the flat-space
  solution~${\varphi = \varphi_b(r_4)}$ into the small
  metric-sensitive part of Eq.~(\ref{eq:flat_action}). This gives,
  \begin{equation}
    \label{eq:8}
    S_b^{\mathrm{BH}} - S_b^{\mathrm{flat}} = \frac{1}{2} \int d^4 y
    \, \delta g^{\mu\nu}   T_{\mu\nu}[\varphi_b]  + O(r_s^{-4})\,,
  \end{equation}
  where~$T_{\mu\nu}$ is an energy-momentum tensor of~$\varphi_b$ and
  the metric change ${\delta g^{\mu\nu} \equiv g_{E}^{\mu\nu} -
    \delta^{\mu\nu}} \propto r_s^{-2}$ is given by
   Eq.~(\ref{eq:7}). It is worth stressing that modification
   of~$\varphi_b$ itself is negligible in Eq.~(\ref{eq:7}), since this
   configuration extremizes the leading flat-space part of~$S_E$. 

  Recall, however, that the flat-space bounce is~$O(4)$ symmetric,
  i.e.\ depends only on  the geodesic distance~$r_4\equiv \sqrt{y^2}$
  to its center: $\varphi_b = \varphi_b(r_4)$. Its  energy-momentum  tensor
  has the form
  \begin{equation*}
    T_{\mu\nu}[\varphi] = (\partial_{r_4} \varphi_b)^2
    \frac{y_\mu y_\nu}{r_4^2}   - g_{\mu\nu} \left[ \frac12
      (\partial_{r_4} \varphi_b)^2 + V(\varphi_b) \right]\,.
  \end{equation*}
  This guarantees that convolution $T_{\mu\nu} \delta g^{\mu\nu}$ with
  the metric perturbation in Eq.~(\ref{eq:7}) is zero, as
  \begin{equation}
    g^{\mu\lambda}{\cal R}^{(0)}_{\mu\nu\lambda\rho}  \equiv {\cal
      R}_{\nu\rho}^{(0)} = 0\quad \mbox{and}\quad y^{\lambda}
    y^{\rho}{\cal R}^{(0)}_{\mu\nu\lambda\rho} = 0
  \end{equation}
  by Einstein equation for Schwarzschild spacetime and symmetry of
  the Riemann tensor.

  We conclude that leading-order gravitational correction in
  Eq.~\eqref{eq:8} equals zero, leaving the next~$O(r_s^{-4})$
  terms. On dimensional grounds,
  \begin{equation}
    \label{eq:9}
    S_b^{\mathrm{BH}}/S_b^{\mathrm{flat}} - 1 = d_b'\, \frac{a^4}{r_s^4}
    \propto O(\gamma_6)\,,
  \end{equation}
  where~$d_b'$ is in general a function of~$ma$.

  Strictly speaking, Eq.~(\ref{eq:9}) is valid only for well-localized
  bounces at~$mr_s \gg 1$. But we will exploit it  at~$mr_s
  \sim O(1)$ and $mr_s \ll 1$ anyway because the respective 
  bounce cores are well localized whereas linear bounce tails
  do not affect the action, cf.\ Eq.~\eqref{eq:S_M_core}. In particular,
  this means that the term~$d_b\sqrt{\gamma_6}$ in the bounce
  action~\eqref{eq:a_R_4} can be calculated in flat spacetime: the 
  metric changes only the next $O(\gamma_6)$ correction in
  Eq.~(\ref{eq:9}).  

  We finish this Appendix with a remark. Euclidean action~$S_E(a)$ in
  Eq.~(\ref{eq:SE_a_g6}) has a single extremum at~${a\sim
    \gamma_6^{1/4}/m}$~--- the minimum representing the bounce. But 
  metric-dependent correction (\ref{eq:9}) with $d_b'<0$ may add a
  maximum at $a\sim mr_s^2$ which resembles our aspherical bubble from
  the second branch. Recall Sec.~\ref{Sec:Aspherical}: these
  second-branch solutions interpolate between the ordinary aspherical
  bubbles and bounces, merging with them at~$r_s = r_s^{(c)}$ and 
  some very small~$r_s$, respectively. At the merging (cusp) points the 
  function~$S_E(a)$ loses one maximum and one minimum. 

%%%%%%%%%%%%%%%%%%%%%%%%%%%%%%%%%%%%%%%%%%%%%%%%

\section{Spherical bubbles and their modes}
\label{App:Modes}

  Here we give details of calculations involving rotationally symmetric
  classical solutions, spherical bubbles and~$O(4)$ invariant
  bounces, and their linear modes.

  We start with the boundary condition for the field on the
  black hole horizon. Recall that Euclidean Sch\-warz\-shild spacetime in
  Fig.~\ref{fig:menzoorka} has the form of a smooth cigar equipped
  with two-spheres at every point, and besides,~${R = R_s}$ is the 
  cigar's tip. We introduce explicitly regular
  coordinates~\eqref{eq:rho_horizon} near the horizon with precise
  radius
  \begin{equation}
    \varrho = 4(R - R_s)\sqrt{R_s / R}\,.
    \label{eq:rho_horizon2}
  \end{equation}
  In this system~$\varrho$ and $\tilde{\tau} \propto \tau$ are polar
  radius and angle near the cigar's tip, whereas $\varrho_\theta \propto
  \theta$ and~$\phi$ are similar polar 
  coordinates on the sphere. Schwarzschild
  interval~(\ref{eq:isotropic_metric}) is regular in these new
  coordinates, 
  \begin{equation*}
    ds_E^2 = \frac{\varrho^2 d\widetilde{\tau}^2}{ \nu_{\varrho}}  +
    \nu_{\varrho} d\varrho^2  + \nu^2_{\varrho} \left(d
    \varrho_{\theta}^2 + 16d \phi^2 R^2_s \sin^2
    \frac{\varrho_{\theta}}{4R_s} \right),
  \end{equation*}
  because it involves only even powers of radii\footnote{This
    property does not hold for the metric~(\ref{eq:isotropic_metric})
    that includes odd and even powers of~${R-R_s}$ in the
    near-horizon expansion. Hence,~$R-R_s$ is not a  smooth polar
    radius.} 
  ~$\varrho$  and~$\varrho_\theta$, where $\nu_{\varrho} = 1 +
  \varrho^2/(64R^2_s)$.

  Now, it is clear that any static ($\tilde{\tau}$-independent) smooth
  configuration~$\varphi (\varrho, \varrho_\theta, \phi)$ satisfies regularity 
  condition~${\prt_{\varrho} \vf  = 0}$ on the horizon~${\varrho =
    0}$. In the original terms this requirement
  gives~${\partial_R \varphi \to 0}$  as~${R \to R_s}$. We impose it
  on spherical~$\varphi_{\mathrm{cb}}^{\mathrm{(s)}}(R)$ and
  aspherical~$\varphi_{\mathrm{cb}}^{\mathrm{(a)}} (R,\theta)$ critical  
  bubbles, as well as on the  static linear
  perturbations~$\xi_{k\ell}(R)$, 
  see the left parts of Eqs.~\eqref{eq:bc_sph} and~\eqref{eq:bc_mod}.
  On the other hand, nonstatic fields $\delta \varphi
  \propto \mathrm{e}^{in\tilde{\tau}}$ depend on the
  angle~$\tilde{\tau}$ and therefore behave as~${\delta 
  \varphi \propto \varrho^{|n|}}$ at~${\varrho \to 0}$. In isotropic
  system, 
  \begin{equation}
    \label{eq:14}
    \delta \varphi \propto (R-R_s)^{|n|} \;\;\, \text{as} \;\;\, R \to R_s
    \quad \mbox{if} \;\;\, \delta \varphi \propto \mathrm{e}^{2\pi in\tau/\beta}.
  \end{equation}
   Perturbations of this kind will be considered below.

  We compute spherical critical bubbles~$
  \varphi_{\mathrm{cb}}^{\mathrm{(s)}}(R)$ numerically  by 
  shooting. Namely, employing the left of Eqs.~\eqref{eq:bc_sph} 
  and~${\varphi_{\mathrm{cb}}^{\mathrm{(s)}}(R_s) = C_\varphi}$ as
  Cauchy data on the horizon, we evolve
  Eq.~\eqref{eq:sph} from~$R = R_s$ to large~$R$ using Bulirsch-Stoer
  method~\cite{Press:2007ipz}. After that we tune~$C_\varphi$ to
  satisfy the falloff condition $\varphi^{\mathrm{(s)}}_{\mathrm{cb}}
  \to 0$ as $R \to  \infty$, see the right part of
  Eq.~\eqref{eq:bc_sph}. Numerical profile of
  $\varphi_{\mathrm{cb}}^{\mathrm{(s)}}(R)$ is shown in  
  Fig.~\ref{fig:BH_cbs}. Euclidean
  actions~$S_E[\varphi_{\mathrm{cb}}^{\mathrm{(s)}}]$ of these
  solutions in Figs.~\ref{fig:intro_answer} and~\ref{fig:full_res} are
  obtained by numerically taking the integral~\eqref{eq:flat_action}
  over~$R$. 
  Flat-space critical bubbles in Fig.~\ref{fig:flat_minphi4} are
  recovered at~${R_s=0}$. 

  Flat-space bounces~$\varphi_b(r_4)$  and static
  modes~$\xi_{k\ell}(R)$ of critical bubbles 
  are provided by the same numerical procedure, albeit with different
  equations to solve. Namely,~$O(4)$ invariant  bounces~$\varphi  = 
  \varphi_b(r_4)$ satisfy flat-space equation
  \begin{equation*}
    \prt_{r_4}^2\varphi_b  + \frac{3}{r_4} 
      \prt_{r_4} \varphi_b- m^2 \varphi_b  + \lambda \varphi^3_b -
      \lambda_6 \varphi_b^5  = 0\,,
  \end{equation*}
  in the regularized model~(\ref{eq:pot}), (\ref{eq:V_reg}) with
  boundary conditions~$\partial_{r_4}\varphi_b = 0$ at~$r_4 = 0$ and $\varphi_b
  \to 0$ at~${r_4 \to  \infty}$. Their actions~$S_E[\varphi_b(r_4)]$
  are shown in Fig.~\ref{fig:S_m_g6} and used in
  Fig.~\ref{fig:horiz_vs_flat}(b). For~$\xi_{k\ell}$, we solve
  Eqs.~\eqref{eq:eigen_combo} at every~$k$ 
  and~$\ell$. In this case~$\xi_{kl}(R_s) = C_{\xi}$ is fixed
  and~$\mu_{k\ell}$ plays the role of shooting  parameter. After
  getting the solution, we extract~$C_\xi$ from the normalization
  condition~\eqref{eq:10}.  This is exactly how the negative
  modes~$\xi_{00}(R)$ and~$\xi_{01}(R)$ were 
  obtained in Sec.~\ref{Sec:Aspherical}, see 
  Eq.~\eqref{eq:spher_deformed} and Fig.~\ref{fig:SE_A0_A1}. 
  The total number of such modes with different~$k$ and~$\ell$ is
  shown in Fig.~\ref{fig:Nmodes_Rs} by circles.

  Now, we derive eigenproblem~(\ref{eq:eigen_combo}) for linear
  modes of spherical bubbles. To this end we write the
  action~(\ref{eq:flat_action}) in isotropic
  coordinates~\eqref{eq:isotropic_metric},
  \begin{multline}
    \label{eq:SE_full}
    S_E = \int d\tau \, dR \, d\theta \, d \phi \,
    \kappa_c	\left[\kappa_k 
      \frac{(\prt_\tau \vf)^2}{2} 
       + \frac{(\prt_R \vf)^2}{2} \right. \\ \left.+ 
      \frac{\left(\partial_{\theta} \vf\right)^2}{2R^2} +
      \frac{\left(\partial_{\phi} \vf\right)^2}{2R^2\sin^2\theta} +
      \frac{r^2}{R^2}  V(\vf)\right]\,,
  \end{multline}
  where~$r(R)$ is given by Eq.\eqref{eq:isotropic_r} and 
  \begin{equation}
    \kappa_c = \left(R^2 - R^2_s\right) \sin \theta\,, \quad 
    \quad
    \kappa_k = \frac{\left(1+R_s/R\right)^{6}}{\left(1-R_s/R\right)^2}
  \end{equation}
  depend on coordinates. Substituting $\varphi =
  \varphi_{\mathrm{cb}}^{\mathrm{(s)}} + \delta \varphi$, we get second
  variation of the action around the solution~$\varphi_{\mathrm{cb}}^{(s)}(R)$, 
  \begin{equation}
    \delta^2 S_E = \frac12 \int d\tau \, dR \, d\theta \, d\phi \; 
    \kappa_c \,  \delta \varphi 
    \hat{L}  \delta \varphi, 
  \end{equation} where \begin{equation} 
    \quad \hat{L} = - \kappa_k \partial^2_\tau - \partial^2_R -
    \frac{2R \partial_R}{R^2 - R^2_s} -
    \frac{\Delta_{\theta\phi}}{R^2}  + \frac{r^2}{R^2}
    V''(\vf^{\mathrm{(s)}}_{\mathrm{cb}}) 
    \label{eq:12}
  \end{equation}
  and $\Delta_{\theta\phi}$ is a spherical Laplacian.

  The next step is to decompose the perturbation $\delta \varphi$ in
  the eigenmodes of operator~$\hat{L}$,
  \begin{equation}
    \delta \varphi = \sum\limits_{k\ell mn}  c_{k\ell mn} \, \xi_{k\ell
      n}(R) \, \mathrm{e}^{ 2\pi in\tau/\beta} Y_{\ell m} (\theta, \phi)\,, 
    \label{eq:tau_expand}
  \end{equation}
  where we already specified dependence on~$\tau$,~$\theta$, and~$\phi$, 
  introduced the coefficients~$c_{k\ell mn}$ and functions~$\xi_{k\ell
    n}(R)$ satisfying the residual eigenproblem
  \begin{equation}
    \hat{L}_{\ell n } \, \xi_{k\ell n} = \left(\hat{L}_{\ell} + \kappa_k 
    n^2 / 64 R^2_s\right) \,\xi_{k\ell n} = \mu_{k\ell n} \,\xi_{kn}
    \label{eq:rad_op_tau}
  \end{equation}
  with eigenvalues~$\mu_{k\ell n}$, where $\hat{L}_\ell$ if given by
  Eq.~\eqref{eq:rad_operator}. Static versions of 
  decomposition~\eqref{eq:tau_expand} and
  eigenproblem~\eqref{eq:rad_op_tau} with~${n=0}$, ${\xi_{k\ell}\equiv
    \xi_{k\ell0}}$, and~$\mu_{k\ell} \equiv \mu_{k\ell 0}$ were used
  in the main text, see  Eqs.~\eqref{eq:sph_decomp}
  and~\eqref{eq:eigen}.

  To explain, why mode expansion is useful, we introduce normalization 
  \begin{equation}
  \label{eq:10}
  \int_{R_s}^{\infty} dR {(R^2 - R_s^2)} \, \xi_{k' \ell n}^* (R) \,
  \xi_{k\ell n}(R) = \delta_{kk'}/\beta
  \end{equation}
  that makes $\hat{L}_{\ell n}$ Hermitian and substitute
  Eq.~\eqref{eq:tau_expand} into Eq.~\eqref{eq:12}. This gives
  Eq.~\eqref{eq:13} with extra sum over~$n$ and thus suggests that 
  mode with $\mu_{k\ell n} < 0$ decreases Euclidean action 
  suppressing vacuum decay. 

\begin{figure}[ht]
  \includegraphics{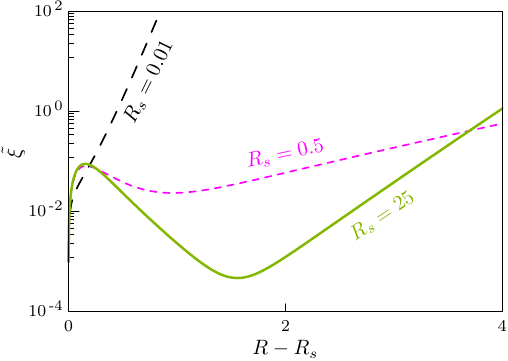}
  \caption{Solutions~$\tilde{\xi}(R)$ of the equation
    ${\hat{L}_{01} \tilde{\xi} = 0}$.}
  \label{fig:tau_modes}
\end{figure}

  We derived the  eigenproblem for Sec.~\ref{Sec:Spher_modes} in
  general~$\tau$-dependent case to demonstrate that
  all negative modes of $\varphi_{\mathrm{cb}}^{\mathrm{(s)}}$ are
  static. Indeed, on the one hand oscillation theorem guarantees that
  the number of  negative modes with~$n=\pm 1$ and~$\ell=0$ equals to
  the number of function~$\tilde{\xi}(R)$ zeros if the latter
  satisfies the  equation~$\hat{L}_{01} \, \tilde{\xi}= 0$ and boundary
  condition~\eqref{eq:14}. Figure~\ref{fig:tau_modes} demonstrates
  that numerical solution for~$\tilde{\xi}(R)$ is 
  sign non-alternating, hence no modes with~$\mu_{k01}<0$.
  On the other hand the operators~$\hat{L}_{\ell n}$ with~${(n,\ell) > (0,1)}$  are strictly
  larger than~$\hat{L}_{01}$ and therefore do not have negative modes either.

  The above calculation shows that arbitrary time-dependent
  perturbations of $\varphi_{\mathrm{cb}}^{\mathrm{(s)}}$ increase
  suppression~$S_E$ of the decay. We conclude that time-dependent
  solutions~--- periodic instantons~\cite{Khlebnikov:1991th,
    Kuznetsov:1997sf} ~-- are irrelevant. 
 	
%%%%%%%%%%%%%%%%%%%%%%%%%%%%%%%%%%%%%%%%%%%%%%%
        
\section{Numerical methods for aspherical bubbles}
\label{App:Numerical}
  Let us describe numerical procedure for computing aspherical
  critical bubbles $\varphi_{\mathrm{cb}}^{\mathrm{(a)}}(R,\theta)$~--- saddle
  points of the action~\eqref{eq:eucl_action_asym}. 

\begin{figure}[ht]
  \includegraphics{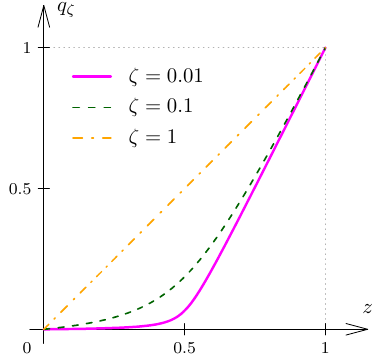}
  \caption{Transfer function $q_\zeta(y)$ in Eq.~\eqref{eq:inhomo_q}
    at different~$\zeta$}.
  \label{fig:trans_fun}
\end{figure}

  We employ dimensionless units with~${R_s=1}$ and two lattices
  depending on whether the bubble is close to the black hole or far
  away from it. In the former case our lattice is nonuniform
  rectangular in~$R$,~$\theta$ coordinates, i.e.\ spherical in the
  physical space. It has $N_R \times N_\theta$ sites 
  $R_j = \exp\left[L_R q_\zeta(w_j)\right]$ and ${\theta_k =
  \pi \,q_\zeta (\vartheta_k)}$ in the region ${0 < R_j <
  \mathrm{e}^{L_R}}$ and ${0 < \theta_j < \pi}$, where ${w_j  \equiv
    (j+\frac12)/N_R}$ and ${\vartheta_k \equiv (k+\frac12)/N_\theta}$
  homogeneously cover the interval ${0 < w_j,\, \vartheta_k <1}$,
  whereas the transfer function 
  \begin{equation}
    \label{eq:inhomo_q}
    q_{\zeta}(y) = y - \frac{1}{2(1-\zeta)} +
    \sqrt{\frac{1}{4(1-\zeta)^2} + y^2 -y }
  \end{equation}
  introduces lattice inhomogeneity controlled
  by~$\zeta$. At ${\zeta=1}$ we have~${q_1(y) = y}$: the lattice is
  uniform in~$\theta$ and exponential in~$R$ with comparable 
  spacings~$\Delta R \propto R$ and~$R\Delta \theta$.  But at 
  $\zeta \ll 1$ a quarter of our lattice is concentrated in the small
  region ${1 < R_j \lesssim \exp(\zeta L_R/2)}$, ${\theta_k \lesssim  
    \pi\zeta/2}$ and the rest of it is almost uniform-exponential,
  see Fig.~\ref{fig:trans_fun}. We use~${\zeta=1}$ for both branches~\footnote{As the secondary branch of aspherical bubbles becomes taller and thinner at small $r_s$, we double-check it using the inhomogeneous grid as well.}
  of aspherical bubbles and~$\zeta = 0.01 \div 0.1$ to resolve cores
  and tails of regularized bounces.

  Our action~\eqref{eq:eucl_action_asym} is discretized using the
  standard second-order replacements, cf.~\cite{Demidov:2011dk, Demidov:2015bua,
    Demidov:2015nea,  Levkov:2017paj, Demidov:2022ljh}. We introduce
  finite differences for derivatives
  $$
  \partial_R \varphi \to  \frac{\varphi_{j+1,k} -
    \varphi_{j,k}}{\Delta R_j}, \qquad 
  \partial_\theta \varphi \to  \frac{\varphi_{j,k+1} -
    \varphi_{j,k}}{\Delta \theta_k}\,,
  $$
  which now belong to the links $(j+1/2,k)$ and $(j, k+1/2)$, as well
  as trapezoidal formulae for the integrals
  $$
    \int dR \, f(R) \to \sum_{j=0}^{N_R-1} f(R_j) \overline{\Delta}
    R_j \quad \mbox{or} \quad \sum_{j=0}^{N_R-2} f_{j+\frac12} \Delta
    R_j \,,
 $$
  where $\Delta R_j \equiv R_{j+1} - R_j$ and
  $\overline{\Delta} R_j \equiv \frac12 (\Delta R_j + \Delta
  R_{j-1})$, $\overline{\Delta} R_0 = \frac12 \Delta R_0$, 
  $\overline{\Delta} R_{N_R-1} = \frac12 \Delta R_{N_R-2}$ are 
  the lattice spacings centered on links and sites, respectively, and 
  $$
  f_{j+\frac12} = \frac12 \left[f(R_{j}) + f(R_{j+1})\right]
  $$
  is the midpoint function value. Integrals of~$f(\theta)$
  over~$\theta$ use similar substitutions involving~$\Delta
  \theta_k$, $\overline{\Delta} \theta_k$, and~$f_{k+\frac12}$.
  We thus get lattice action 
    \begin{align*}
     \frac{S_E}{32\pi^2} = &  \sum\limits_{j=0}^{N_R-2} \sum\limits_{k=0}^{N_\theta-1}
      \frac{\left(\varphi_{j+1,k}-\varphi_{j,k}\right)^2}{2 \, \Delta
        R_j} \, (\kappa_{c})_{j+\frac12, k}\,
      \overline{\Delta}\theta_k\\
      + & \sum\limits_{j=0}^{N_R-1} \sum\limits_{k=0}^{N_\theta-2}
    \frac{\left(\varphi_{j,k+1}-\varphi_{j,k}\right)^2}{2 \,
      \Delta\theta_k}   \frac{(\kappa_c)_{j,
        k+ \frac12}}{R^2_j}\, \overline{\Delta} R_j\\
     + & \sum\limits_{j=0}^{N_R-1} \sum\limits_{k=0}^{N_\theta-1}
    (\kappa_c)_{j,k} \, \frac{r_{j}^2}{R_{j}^2}\, V(\varphi_{j,k}) \,
    \overline{\Delta} R_j \overline{\Delta}\theta_k\,,
    \label{eq:discr_pot}
  \end{align*}
  where $\kappa_c = (R^2 - R_s^2) \sin\theta$, see
  Eq.~\eqref{eq:eucl_action_asym}. 

  On the lattice, critical bubbles extremize discretized action~$S_E$, 
  i.e.\ satisfy a system of~${N_R \times N_\theta}$ algebraic
  equations~${\prt S_{E} / \prt \vf_{jk} = 0}$ for the same number of  
  unknowns~$\varphi_{j,k}$. We solve the latter by
  Newton--Raphson iterations~\cite{Demidov:2011dk, 
    Demidov:2015bua, Demidov:2015nea,  Levkov:2017paj,
    Demidov:2022ljh, Press:2007ipz}. Namely, starting from the
  approximation~$\varphi_{j,k}^{(0)}$, we repeatedly compute
  correction~${\delta \varphi = \varphi - \varphi^{(0)}}$ to it using
  the linearized system
  \begin{equation}
    \sum_{j'k'}\, \frac{\partial^2 S_{E}}{\partial
      \varphi_{j,k}\partial \varphi_{j'\!,k'}} \;\delta \varphi_{j'\!,k'} = -
    \frac{\partial S_{E}}{\partial \varphi_{j,k}}\,.
    \label{eq:lin_lat}
  \end{equation}
  After that we update the approximation $\varphi^{(0)} \to
  \varphi^{(0)} + \delta \varphi$ and start new iteration. The
  iterations stop when the nonlinear equations are solved to a desired
  tolerance: $\max \left|\prt S_{E} / \prt \vf_{j,k}\right|<
 10^{-9}$. Notably, convergence of this method requires good initial
  approximation~$\varphi^{(0)}$ which is  therefore chosen carefully
  in Secs.~\ref{Sec:Aspherical}, \ref{Sec:Lipatons}. We solve the linear 
  system~\eqref{eq:lin_lat} by minimal residual~\cite{Press:2007ipz}
  modification of linear conjugate gradient
  method~\cite{shewchuk1994introduction}.  

  In practical calculations we set~${L_R = 5.5}$, ${N_R  = 700}$, 
  and ${N_\theta  = 400}$ and check that the Newton--Raphson method is
  converged to~$\delta \varphi/\varphi \lesssim
  10^{-6}$. Discretization and finite-volume effects are controlled by
  changing~$N_R$, $N_\theta$, and~$L_R$. This reveals relative
  errors of our solutions and their suppressions~$S_E$ to be of 
  order~$10^{-3}$ on the second aspherical branch of critical bubbles
  and of order~$10^{-4}$ on the main branch and for small-size  
  bounces.   

  If the bubble moves far away from the black hole due to~$Z_c$ fixation
  in Eq.~\eqref{eq:S-lC}, we seamlessly
  switch to~${N_z\times N_\sigma}$ cylindrical
  grid~${(z_j,\ \theta_k)}$ in coordinates 
  \begin{equation}
    z = R \cos \theta \qquad \mbox{and} \qquad  \sigma = R \sin \theta\,.
    \label{eq:cylindrical}
  \end{equation}
  The sites $z_j = Z_c +L_z (2j - N_z +1)/2N_z$ and ${\sigma_k = L_\sigma 
  (k + \frac12)/N_\sigma}$ of this new lattice uniformly cover the
  region $Z_c - \frac12 L_z < z < Z_c + \frac12 L_z$ and $\sigma <
  L_\sigma$, where~$Z_c$ is the bubble central position in
  Eq.~(\ref{eq:Z_d}). Respective discretized action~$S_E$ is found 
  by expressing Eq.~\eqref{eq:eucl_action_asym} in terms of~$z$
  and~$\sigma$ and using the standard finite-difference expressions for
  the derivatives and integrals. Otherwise, the above numerical
  procedure applies without change to~$S_E$ and the modified
  functional~\eqref{eq:S-lC}. In practice, we use $N_z \times
  N_\sigma= 400\times 200$ lattice and cutoffs~$L_z = 20$, $L_{\sigma}
  = 10$ thus stabilizing all relative errors at the level
    of~$10^{-3}$.

  It is worth noting that flat-space periodic
  instantons~$\varphi(\tau, |\mbs{x}|)$ in~Fig.~\ref{fig:flat_minphi4}
  are computed using the same method and uniform lattice in~$\tau$
  and~$|\mbs{x}|$.

%%%%%%%%%%%%%%%%%%%%%%%%%%%%%%%%%%%%%%%%%%%%%%%%%
	
	\bibliography{bh_inst}{}
	\end{document}